%% file: draft10T.tex
\documentclass[epj,nopacs]{svjoura}
\usepackage{amsmath}
\usepackage{graphics}
\usepackage{graphicx}
\usepackage{epsfig}
\usepackage{amssymb}
\usepackage{units}
\usepackage{cite}
\usepackage{lineno}
\usepackage{calc}
\usepackage{endnotes}
\usepackage{fixltx2e}
\usepackage{float}
\usepackage{multirow}

\newcommand{\dndeta}[1]{${\rm d}N_{\rm ch}/{\rm d}\eta = #1$}
\newcommand{\dNdEta}{{\rm d}N_{\rm ch}/{\rm d}\eta }
\newcommand{\bfig}[1][t]{\begin{figure}[#1] \begin{center}}
\newcommand{\efig}{\end{center} \end{figure}}

\newcommand{\btab}[1][ht!]{\begin{table}[#1] \begin{center}}
\newcommand{\etab}{\end{center} \end{table}}

\newcommand{\bq}{\begin{equation}}
\newcommand{\eq}{\end{equation}}
\newcommand{\tblspc}{\rule{0pt}{2.5ex}}
\newcommand{\pbar}{\bar{\mbox{p}}}
\newcommand{\spps}{Sp$\pbar$S }
\newcommand{\etain}[1]{$|\eta|<#1$}
\newcommand{\cms}{\sqrt{s}}
\newcounter{vers}

\usepackage{grffile}
\usepackage{preprintcover}
\PreprintIdNumber{}
\PreprintCoverPaperTitle{Charged-particle multiplicity measurement in proton-proton
collisions at ${\sqrt{{s}}= 0.9}$ and $2.36$\,TeV with ALICE at LHC}
\PreprintCoverAbstract{Charged-particle production was studied in proton--proton collisions collected at the LHC with the ALICE detector at centre-of-mass energies $0.9$~TeV and 2.36 TeV in the pseudorapidity range \etain{1.4}. In the central region (\etain{0.5}), at $0.9$~TeV, we measure charged-particle pseudorapidity density $\dNdEta = 3.02 \pm 0.01 (\emph{stat.}) ^{+0.08}_{-0.05} (\emph{syst.})$ for inelastic interactions, and $\dNdEta = 3.58 \pm 0.01(\emph{stat.}) ^{+0.12}_{-0.12} (\emph{syst.})$ for non-single-diffractive interactions. At 2.36~TeV, we find $\dNdEta = 3.77 \pm 0.01(\emph{stat.}) ^{+0.25}_{-0.12} (\emph{syst.})$ for inelastic, and $\dNdEta = 4.43 \pm 0.01(\emph{stat.}) ^{+0.17}_{-0.12} (\emph{syst.})$ for non-single-diffractive collisions. The relative increase in charged-particle multiplicity from the lower to higher energy is $24.7\% \pm 0.5\%(\emph{stat.}) ^{+5.7}_{-2.8}\%(\emph{syst.})$  for inelastic and  $23.7\% \pm 0.5\%(\emph{stat.}) ^{+4.6}_{-1.1}\%(\emph{syst.})$ for non-single-diffractive interactions. This increase is consistent with that reported by the CMS collaboration for non-single-diffractive events and larger than that found by a number of commonly used models. The multiplicity distribution was measured in different pseudorapidity intervals and studied in terms of KNO variables at both energies. The results are compared to proton--antiproton data and to model predictions.}
\PreprintJournalName{EPJC}

\begin{document}
\hugehead
\title{Charged-particle multiplicity measurement in proton-proton
collisions at ${\sqrt{{\it s}}= 0.9}$ and ${2.36}$\,TeV with ALICE at LHC}
\subtitle{ALICE collaboration}
\input{authors10T.tex}
\input{institutes7.tex}
%
\date{}
%
\input{abstractpaper2.tex}

\maketitle
%
%

\input{corpuspaper2.tex}

\input{acknowledgements3.tex}

\input{bibliographypaper2.tex}
\end{document}

%% file: authors10T.tex
\author{  
This publication is dedicated to the memory of our colleague Hans-{\AA}ke~Gustafsson.
\\
\\
K.~Aamodt\inst{79} \and  
N.~Abel\inst{42} \and  
U.~Abeysekara\inst{77} \and  
A.~Abrahantes~Quintana\inst{41} \and  
A.~Abramyan\inst{114} \and  
D.~Adamov\'{a}\inst{87} \and  
M.M.~Aggarwal\inst{25} \and  
G.~Aglieri~Rinella\inst{39} \and  
A.G.~Agocs\inst{18} \and  
S.~Aguilar~Salazar\inst{65}
% ...
\renewcommand{\notesname}{Affiliation notes}  
\endnotetext[1]{Also at{ Laboratoire de Physique Corpusculaire (LPC), Clermont Universit\'{e}, Universit\'{e} Blaise Pascal, CNRS--IN2P3, Clermont-Ferrand, France}}  
\endnotetext[2]{Also at{ Frankfurt Institute for Advanced Studies, Johann Wolfgang Goethe-Universit\"{a}t Frankfurt, Frankfurt, Germany}}  
\endnotetext[3]{Now at{ Sezione INFN, Padova, Italy}}  
\endnotetext[4]{Now at{ Research Division and ExtreMe Matter Institute EMMI, GSI Helmholtzzentrum f\"{u}r Schwerionenforschung, Darmstadt, Germany}}  
\endnotetext[5]{Now at{ Institut f\"{u}r Kernphysik, Johann Wolfgang Goethe-Universit\"{a}t Frankfurt, Frankfurt, Germany}}  
\endnotetext[6]{Now at{ Physics Department, University of Cape Town, iThemba Laboratories, Cape Town, South Africa}}  
\endnotetext[7]{Now at{ National Institute for Physics and Nuclear Engineering, Bucharest, Romania}}  
\endnotetext[8]{Also at{ University of Houston, Houston, TX, United States}}  
\endnotetext[9]{Now at{ Faculty of Science, P.J.~\v{S}af\'{a}rik University, Ko\v{s}ice, Slovakia}}  
\endnotetext[10]{Now at{ European Organization for Nuclear Research (CERN), Geneva, Switzerland}}  
\endnotetext[11]{Now at{ Helsinki Institute of Physics (HIP) and University of Jyv\"{a}skyl\"{a}, Jyv\"{a}skyl\"{a}, Finland}}  
\endnotetext[12]{Now at{ Institut Pluridisciplinaire Hubert Curien (IPHC), Universit\'{e} de Strasbourg, CNRS-IN2P3, Strasbourg, France}}  
\endnotetext[13]{Now at{ Sezione INFN, Bari, Italy}}  
\endnotetext[14]{Now at{ Institut f\"{u}r Kernphysik, Westf\"{a}lische Wilhelms-Universit\"{a}t M\"{u}nster, M\"{u}nster, Germany}}  
\endnotetext[15]{Now at: University of Technology and Austrian Academy of Sciences, Vienna, Austria}  
\endnotetext[16]{Also at{ Lawrence Livermore National Laboratory, Livermore, CA, United States}}  
\endnotetext[17]{Also at{ European Organization for Nuclear Research (CERN), Geneva, Switzerland}}  
\endnotetext[18]{Now at { Secci\'{o}n F\'{\i}sica, Departamento de Ciencias, Pontificia Universidad Cat\'{o}lica del Per\'{u}, Lima, Peru}}  
\endnotetext[19]{Deceased}  
\endnotetext[20]{Now at{ Yale University, New Haven, CT, United States}}  
\endnotetext[21]{Now at{ University of Tsukuba, Tsukuba, Japan}}  
\endnotetext[22]{Also at { Centro Fermi -- Centro Studi e Ricerche e Museo Storico della Fisica ``Enrico Fermi'', Rome, Italy}}  
\endnotetext[23]{Now at{ Dipartimento Interateneo di Fisica `M.~Merlin' and Sezione INFN, Bari, Italy}}  
\endnotetext[24]{Also at{ Laboratoire de Physique Subatomique et de Cosmologie (LPSC), Universit\'{e} Joseph Fourier, CNRS-IN2P3, Institut Polytechnique de Grenoble, Grenoble, France}}  
\endnotetext[25]{Now at{ Dipartimento di Fisica Sperimentale dell'Universit\`{a} and Sezione INFN, Turin, Italy}}  
\endnotetext[26]{Now at{ Physics Department, Creighton University, Omaha, NE, United States}}  
\endnotetext[27]{Now at{ Commissariat \`{a} l'Energie Atomique, IRFU, Saclay, France}}  
\endnotetext[28]{Also at{ Department of Physics, University of Oslo, Oslo, Norway}}  
\endnotetext[29]{Now at{ Physikalisches Institut, Ruprecht-Karls-Universit\"{a}t Heidelberg, Heidelberg, Germany}}  
\endnotetext[30]{Now at{ Institut f\"{u}r Kernphysik, Technische Universit\"{a}t Darmstadt, Darmstadt, Germany}}  
\endnotetext[31]{Now at{ Department of Physics and Technology, University of Bergen, Bergen, Norway}}  
\endnotetext[32]{Now at{ Physics Department, University of Athens, Athens, Greece}}  
\endnotetext[33]{Also at{ Institut f\"{u}r Kernphysik, Westf\"{a}lische Wilhelms-Universit\"{a}t M\"{u}nster, M\"{u}nster, Germany}}  
\endnotetext[34]{Now at{ SUBATECH, Ecole des Mines de Nantes, Universit\'{e} de Nantes, CNRS-IN2P3, Nantes, France}}  
\endnotetext[35]{Now at{ Universit\'{e} de Lyon 1, CNRS/IN2P3, Institut de Physique Nucl\'{e}aire de Lyon, Lyon, France}}  
\endnotetext[36]{Now at: Centre de Calcul IN2P3, Lyon, France}  
\endnotetext[37]{Now at{ Variable Energy Cyclotron Centre, Kolkata, India}}  
\endnotetext[38]{Also at{ Dipartimento di Fisica dell'Universit\`{a} and Sezione INFN, Padova, Italy}}  
\endnotetext[39]{Also at{ Sezione INFN, Bologna, Italy}}  
\endnotetext[40]{Also at Dipartimento di Fisica dell\'{ }Universit\`{a}, Udine, Italy}  
\endnotetext[41]{Also at Wroc{\l}aw University, Wroc{\l}aw, Poland} 
\bigskip  
\theendnotes  
\section*{Collaboration institutes}  
}  

%% file: institutes7.tex
\institute{ 
Dipartimento di Scienze e Tecnologie Avanzate dell'Universit\`{a} del Piemonte Orientale and Gruppo Collegato INFN, Alessandria, Italy 
\and 
Department of Physics Aligarh Muslim University, Aligarh, India 
\and 
National Institute for Nuclear and High Energy Physics (NIKHEF), Amsterdam, Netherlands 
\and 
Physics Department, University of Athens, Athens, Greece 
\and 
Dipartimento Interateneo di Fisica `M.~Merlin' and Sezione INFN, Bari, Italy 
\and 
Sezione INFN, Bari, Italy 
\and 
China Institute of Atomic Energy, Beijing, China 
\and 
Department of Physics and Technology, University of Bergen, Bergen, Norway 
\and 
Faculty of Engineering, Bergen University College, Bergen, Norway 
\and 
Lawrence Berkeley National Laboratory, Berkeley, CA, United States 
\and 
Institute of Physics, Bhubaneswar, India 
\and 
School of Physics and Astronomy, University of Birmingham, Birmingham, United Kingdom 
\and 
Dipartimento di Fisica dell'Universit\`{a} and Sezione INFN, Bologna, Italy 
\and 
Sezione INFN, Bologna, Italy 
\and 
Faculty of Mathematics, Physics and Informatics, Comenius University, Bratislava, Slovakia 
\and 
Institute of Space Sciences (ISS), Bucharest, Romania 
\and 
National Institute for Physics and Nuclear Engineering, Bucharest, Romania 
\and 
KFKI Research Institute for Particle and Nuclear Physics, Hungarian Academy of Sciences, Budapest, Hungary 
\and 
Dipartimento di Fisica dell'Universit\`{a} and Sezione INFN, Cagliari, Italy 
\and 
Sezione INFN, Cagliari, Italy 
\and 
Universidade Estadual de Campinas (UNICAMP), Campinas, Brazil 
\and 
Physics Department, University of Cape Town, iThemba Laboratories, Cape Town, South Africa 
\and 
Dipartimento di Fisica e Astronomia dell'Universit\`{a} and Sezione INFN, Catania, Italy 
\and 
Sezione INFN, Catania, Italy 
\and 
Physics Department, Panjab University, Chandigarh, India 
\and 
Laboratoire de Physique Corpusculaire (LPC), Clermont Universit\'{e}, Universit\'{e} Blaise Pascal, CNRS--IN2P3, Clermont-Ferrand, France 
\and 
Department of Physics, Ohio State University, Columbus, OH, United States 
\and 
Niels Bohr Institute, University of Copenhagen, Copenhagen, Denmark 
\and 
The Henryk Niewodniczanski Institute of Nuclear Physics, Polish Academy of Sciences, Cracow, Poland 
\and 
Universidad Aut\'{o}noma de Sinaloa, Culiac\'{a}n, Mexico 
\and 
Research Division and ExtreMe Matter Institute EMMI, GSI Helmholtzzentrum f\"{u}r Schwerionenforschung, Darmstadt, Germany 
\and 
Institut f\"{u}r Kernphysik, Technische Universit\"{a}t Darmstadt, Darmstadt, Germany 
\and 
Wayne State University, Detroit, MI, United States 
\and 
Joint Institute for Nuclear Research (JINR), Dubna, Russia 
\and 
Frankfurt Institute for Advanced Studies, Johann Wolfgang Goethe-Universit\"{a}t Frankfurt, Frankfurt, Germany 
\and 
Institut f\"{u}r Kernphysik, Johann Wolfgang Goethe-Universit\"{a}t Frankfurt, Frankfurt, Germany 
\and 
Laboratori Nazionali di Frascati, INFN, Frascati, Italy 
\and 
Petersburg Nuclear Physics Institute, Gatchina, Russia 
\and 
European Organization for Nuclear Research (CERN), Geneva, Switzerland 
\and 
Laboratoire de Physique Subatomique et de Cosmologie (LPSC), Universit\'{e} Joseph Fourier, CNRS-IN2P3, Institut Polytechnique de Grenoble, Grenoble, France 
\and 
Centro de Aplicaciones Tecnol\'{o}gicas y Desarrollo Nuclear (CEADEN), Havana, Cuba 
\and 
Kirchhoff-Institut f\"{u}r Physik, Ruprecht-Karls-Universit\"{a}t Heidelberg, Heidelberg, Germany 
\and 
Physikalisches Institut, Ruprecht-Karls-Universit\"{a}t Heidelberg, Heidelberg, Germany 
\and 
Hiroshima University, Hiroshima, Japan 
\and 
University of Houston, Houston, TX, United States 
\and 
Physics Department, University of Rajasthan, Jaipur, India 
\and 
Physics Department, University of Jammu, Jammu, India 
\and 
Helsinki Institute of Physics (HIP) and University of Jyv\"{a}skyl\"{a}, Jyv\"{a}skyl\"{a}, Finland 
\and 
Gangneung-Wonju National University, Gangneung, South Korea 
\and 
Scientific Research Technological Institute of Instrument Engineering, Kharkov, Ukraine 
\and 
Bogolyubov Institute for Theoretical Physics, Kiev, Ukraine 
\and 
University of Tennessee, Knoxville, TN, United States 
\and 
Saha Institute of Nuclear Physics, Kolkata, India 
\and 
Variable Energy Cyclotron Centre, Kolkata, India 
\and 
Fachhochschule K\"{o}ln, K\"{o}ln, Germany 
\and 
Faculty of Science, P.J.~\v{S}af\'{a}rik University, Ko\v{s}ice, Slovakia 
\and 
Institute of Experimental Physics, Slovak Academy of Sciences, Ko\v{s}ice, Slovakia 
\and 
Laboratori Nazionali di Legnaro, INFN, Legnaro, Italy 
\and 
Secci\'{o}n F\'{\i}sica, Departamento de Ciencias, Pontificia Universidad Cat\'{o}lica del Per\'{u}, Lima, Peru 
\and 
Lawrence Livermore National Laboratory, Livermore, CA, United States 
\and 
Division of Experimental High Energy Physics, University of Lund, Lund, Sweden 
\and 
Universit\'{e} de Lyon 1, CNRS/IN2P3, Institut de Physique Nucl\'{e}aire de Lyon, Lyon, France 
\and 
Centro de Investigaciones Energ\'{e}ticas Medioambientales y Tecnol\'{o}gicas (CIEMAT), Madrid, Spain 
\and 
Instituto de Ciencias Nucleares, Universidad Nacional Aut\'{o}noma de M\'{e}xico, Mexico City, Mexico 
\and 
Instituto de F\'{\i}sica, Universidad Nacional Aut\'{o}noma de M\'{e}xico, Mexico City, Mexico 
\and 
Centro de Investigaci\'{o}n y de Estudios Avanzados (CINVESTAV), Mexico City and M\'{e}rida, Mexico 
\and 
Institute for Nuclear Research, Academy of Sciences, Moscow, Russia 
\and 
Institute for Theoretical and Experimental Physics, Moscow, Russia 
\and 
Moscow Engineering Physics Institute, Moscow, Russia 
\and 
Russian Research Centre Kurchatov Institute, Moscow, Russia 
\and 
Indian Institute of Technology, Mumbai, India 
\and 
Institut f\"{u}r Kernphysik, Westf\"{a}lische Wilhelms-Universit\"{a}t M\"{u}nster, M\"{u}nster, Germany 
\and 
SUBATECH, Ecole des Mines de Nantes, Universit\'{e} de Nantes, CNRS-IN2P3, Nantes, France 
\and 
Yale University, New Haven, CT, United States 
\and 
Budker Institute for Nuclear Physics, Novosibirsk, Russia 
\and 
Oak Ridge National Laboratory, Oak Ridge, TN, United States 
\and 
Physics Department, Creighton University, Omaha, NE, United States 
\and 
Institut de Physique Nucl\'{e}aire d'Orsay (IPNO), Universit\'{e} Paris-Sud, CNRS-IN2P3, Orsay, France 
\and 
Department of Physics, University of Oslo, Oslo, Norway 
\and 
Dipartimento di Fisica dell'Universit\`{a} and Sezione INFN, Padova, Italy 
\and 
Sezione INFN, Padova, Italy 
\and 
Faculty of Nuclear Sciences and Physical Engineering, Czech Technical University in Prague, Prague, Czech Republic 
\and 
Institute of Physics, Academy of Sciences of the Czech Republic, Prague, Czech Republic 
\and 
Institute for High Energy Physics, Protvino, Russia 
\and 
Benem\'{e}rita Universidad Aut\'{o}noma de Puebla, Puebla, Mexico 
\and 
Pusan National University, Pusan, South Korea 
\and 
Nuclear Physics Institute, Academy of Sciences of the Czech Republic, \v{R}e\v{z} u Prahy, Czech Republic 
\and 
Dipartimento di Fisica dell'Universit\`{a} `La Sapienza' and Sezione INFN, Rome, Italy 
\and 
Sezione INFN, Rome, Italy 
\and 
Centro Fermi -- Centro Studi e Ricerche e Museo Storico della Fisica ``Enrico Fermi'', Rome, Italy 
\and 
Commissariat \`{a} l'Energie Atomique, IRFU, Saclay, France 
\and 
Dipartimento di Fisica `E.R.~Caianiello' dell'Universit\`{a} and Sezione INFN, Salerno, Italy 
\and 
California Polytechnic State University, San Luis Obispo, CA, United States 
\and 
Departamento de F\'{\i}sica de Part\'{\i}culas and IGFAE, Universidad de Santiago de Compostela, Santiago de Compostela, Spain 
\and 
Universidade de S\~{a}o Paulo (USP), S\~{a}o Paulo, Brazil 
\and 
Russian Federal Nuclear Center (VNIIEF), Sarov, Russia 
\and 
Department of Physics, Sejong University, Seoul, South Korea 
\and 
Yonsei University, Seoul, South Korea 
\and 
Technical University of Split FESB, Split, Croatia 
\and 
V.~Fock Institute for Physics, St. Petersburg State University, St. Petersburg, Russia 
\and 
Institut Pluridisciplinaire Hubert Curien (IPHC), Universit\'{e} de Strasbourg, CNRS-IN2P3, Strasbourg, France 
\and 
University of Tokyo, Tokyo, Japan 
\and 
Dipartimento di Fisica dell'Universit\`{a} and Sezione INFN, Trieste, Italy 
\and 
Sezione INFN, Trieste, Italy 
\and 
University of Tsukuba, Tsukuba, Japan 
\and 
Dipartimento di Fisica Sperimentale dell'Universit\`{a} and Sezione INFN, Turin, Italy 
\and 
Sezione INFN, Turin, Italy 
\and 
Institute for Subatomic Physics, Utrecht University, Utrecht, Netherlands 
\and 
Soltan Institute for Nuclear Studies, Warsaw, Poland 
\and 
Warsaw University of Technology, Warsaw, Poland 
\and 
Purdue University, West Lafayette, IN, United States 
\and 
Zentrum f\"{u}r Technologietransfer und Telekommunikation (ZTT), Fachhochschule Worms, Worms, Germany 
\and 
Hua-Zhong Normal University, Wuhan, China 
\and 
Yerevan Physics Institute, Yerevan, Armenia 
\and 
Rudjer Bo\v{s}kovi\'{c} Institute, Zagreb, Croatia 
} 

%% file: abstractpaper2.tex
\abstract{Charged-particle production was studied in proton--proton collisions collected at the LHC with the ALICE detector at centre-of-mass energies $0.9$~TeV and 2.36 TeV in the pseudorapidity range \etain{1.4}. In the central region (\etain{0.5}), at $0.9$~TeV, we measure charged-particle pseudorapidity density $\dNdEta = 3.02 \pm 0.01 (\emph{stat.}) ^{+0.08}_{-0.05} (\emph{syst.})$ for inelastic interactions, and $\dNdEta = 3.58 \pm 0.01(\emph{stat.}) ^{+0.12}_{-0.12} (\emph{syst.})$ for non-single-diffractive interactions. At 2.36~TeV, we find $\dNdEta = 3.77 \pm 0.01(\emph{stat.}) ^{+0.25}_{-0.12} (\emph{syst.})$ for inelastic, and $\dNdEta = 4.43 \pm 0.01(\emph{stat.}) ^{+0.17}_{-0.12} (\emph{syst.})$ for non-single-diffractive collisions. The relative increase in charged-particle multiplicity from the lower to higher energy is $24.7\% \pm 0.5\%(\emph{stat.}) ^{+5.7}_{-2.8}\%(\emph{syst.})$  for inelastic and  $23.7\% \pm 0.5\%(\emph{stat.}) ^{+4.6}_{-1.1}\%(\emph{syst.})$ for non-single-diffractive interactions. This increase is consistent with that reported by the CMS collaboration for non-single-diffractive events and larger than that found by a number of commonly used models. The multiplicity distribution was measured in different pseudorapidity intervals and studied in terms of KNO variables at both energies. The results are compared to proton--antiproton data and to model predictions.}

%% file: corpuspaper2.tex
\section{Introduction}

Whenever entering a new energy regime with hadron colliders, it is important to measure the global characteristics  of the collisions.
These interactions, dominated by soft (i.e. small-momentum-transfer) processes, are useful to study QCD in the non-perturbative regime, and to constrain phenomenological models and event generators.
Such studies are also important for the understanding of backgrounds for measurements of hard and rare interactions.

ALICE~\cite{ALICEdet} has measured the pseudorapidity density of charged particles produced in proton--proton collisions at a centre-of-mass energy $\cms = 900$~GeV~\cite{ALICEfirst} with low statistics from the first collisions at the CERN Large Hadron Collider (LHC)~\cite{LHC}. Results were given for two normalizations:
\begin{itemize}
 \item{inelastic (INEL); this corresponds to the sum of all inelastic interactions (non-diffractive ND, single-diffrac\-tive SD, and double-diffractive DD) with the trigger biases corrected for each event class individually according to their respective estimated abundances and trigger efficiencies;}
 \item{non-single-diffractive (NSD); here the corrections are applied to non-diffractive and double-diffractive processes only, while removing, on average, the single-diffractive contribution.}
\end{itemize}
The corrections to INEL and NSD samples are based on previous experimental data and simulations with Monte Carlo event generators. Charged-particle pseudorapidity density in pp collisions at LHC was also published by the CMS collaboration for NSD interactions~\cite{CMS_first}, and by the ATLAS collaboration for a different event selection~\cite{ATLAS_first}, not directly comparable with our measurements and those of CMS.

We have used the first high energy proton--proton collisions at the LHC at a centre-of-mass energy $\cms =$ \linebreak $2.36$~TeV, as well as a larger statistics data sample at $\cms = \unit[0.9]{TeV}$, to determine the pseudorapidity density of charged-pri\-ma\-ry particles\footnote{Primary particles are defined as prompt particles produced in the collision and all decay products, except products from weak decays of strange particles. }, $\dNdEta$, in the central pseudorapidity region (\etain{1.4}). According to commonly used models \cite{QGSMcal,Pythia,Pythia1,D6Ttune,CSCtune,Perugiatune,PhoJet},
an increase in $\dNdEta$ of 17--22\,\% for INEL events and of 14--19\,\% for NSD events is expected in 2.36~TeV collisions relative to 0.9~TeV collisions.

We also studied the distribution of the multiplicity of charged particles in the central pseudorapidity region (\etain{1.3}). The multiplicity distribution of charged particles (the probability $P(N_{\mathrm{ch}})$ that a collision has multiplicity $N_{\mathrm{ch}}$) can be described by KNO scaling~\cite{kno_scaling} over a wide energy range. KNO scaling means that the distribution $\langle N_{\mathrm{ch}} \rangle P(z)$, where $z = N_{\mathrm{ch}} / \langle N_{\mathrm{ch}} \rangle$, is independent of energy.
In full phase space, scaling holds up to the top ISR energy (pp at $\cms = \unit[62.2]{GeV}$) \cite{mult_isr}. Deviations from scaling are observed at higher energies, starting at \unit[200]{GeV} with p$\pbar$ collisions at the \spps collider \cite{ua5_mult1}. However, in limited central $\eta$-intervals scaling has been found to hold up to \unit[900]{GeV}. The UA5 collaboration~\cite{ua5_mult2} observed scaling for non-single-diffractive events in restricted central $\eta$-intervals
and its progressive violation with increasing $\eta$-ranges. The UA1 collaboration~\cite{ua1_mult} also observed scaling in a larger interval \etain{2.5}.
In inelastic events, deviation from KNO scaling was observed in full phase space already at ISR energies \cite{mult_isr}.
Such deviations are generally attributed to semi-hard gluon radiation (minijets) and to multi-parton scattering.

The Negative-Binomial Distribution (NBD) \cite{binom} describes multiplicity distributions in full phase space up to \unit[540]{GeV}; however, this description is not successful at \unit[900]{GeV}~\cite{ua5_mult3}. NBD describes the distributions up to \unit[1.8]{TeV} in limited $\eta$-intervals (\etain{0.5})~\cite{cdf_multiplicity1}. For larger $\eta$-intervals and in full phase space, only the sum of two NBDs provides a reasonable fit~\cite{twocomponent, twocomponent_etarange}.

Comparing these multiplicity measurements with the predictions of Monte Carlo generators used by the LHC experiments will allow a better tuning of these models to accurately simulate minimum-bias and underlying-event effects. 
A recent review of multiplicity measurements at high energies can be found in~\cite{Grosse}.

This article is organized as follows: a description of the ALICE detector subsystems used in this analysis is presented in Section 2; Section 3 is dedicated to the definition of the event samples; Section 4 to data analysis; in Section 5 systematic uncertainties are discussed; the results are given in Section 6 and Section 7 contains the conclusions.

\begin{figure}[b!]
\centering
\includegraphics[width=\linewidth]{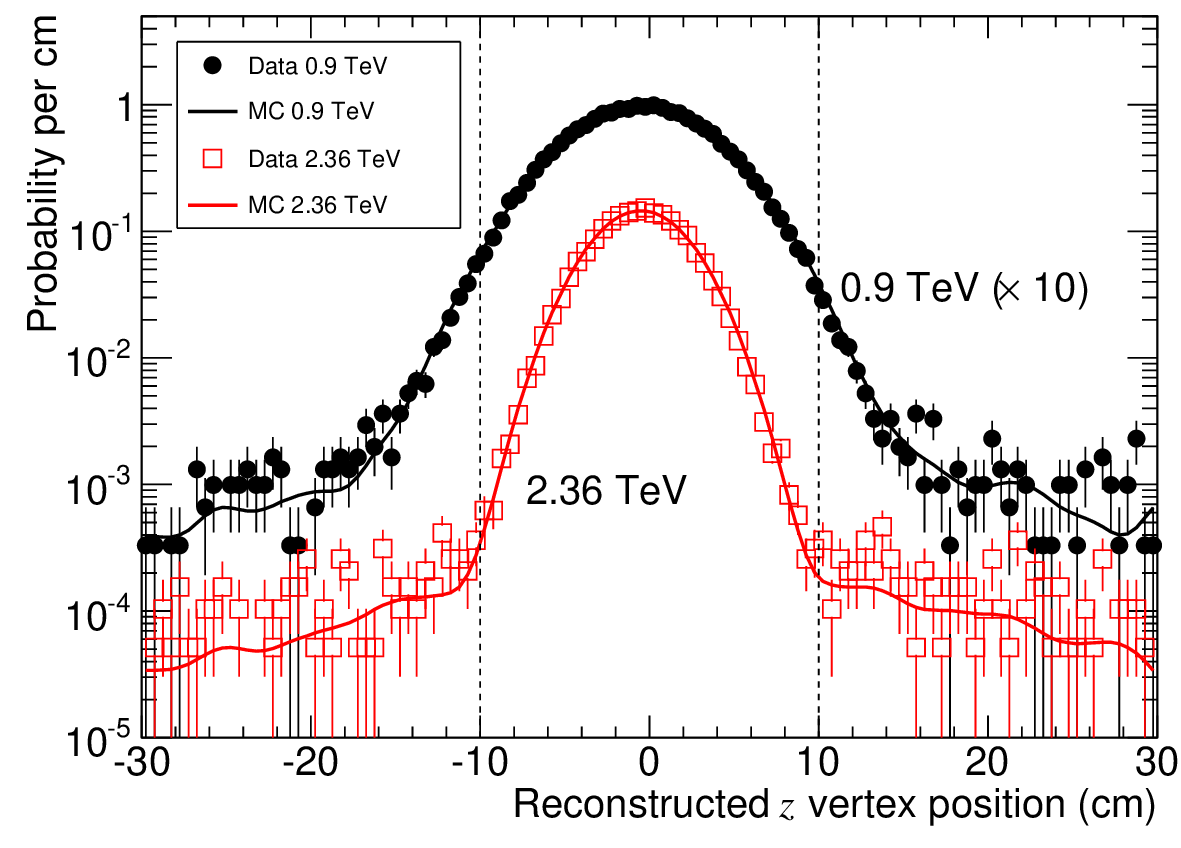}
\caption{Distributions of reconstructed event vertices along the beam direction ($z$) obtained from hit correlations in the two pixel layers of the ALICE inner tracking system for the event samples used in the analysis (see text): $\cms = \unit[0.9]{TeV}$ (full symbols) and $\cms = \unit[2.36]{TeV}$ (open symbols). The lines are from Monte Carlo simulations.
Vertical dashed lines delimit the region $|z|<10$~cm, where the events for the present analysis were selected.}
\label{figVtx}
\end{figure}

\section{The ALICE experiment and data collection}

The ALICE experiment consists of a large number
of detector subsystems which are described in detail in~\cite{ALICEdet}. This analysis is based mainly on data from the Silicon Pixel Detector (SPD), since it has the largest pseudorapidity coverage in the central region and is located closest to the interaction region, implying a very low momentum cut-off and a small contamination from secondary particles.

The SPD detector surrounds the central beryllium \linebreak beam pipe (3 cm radius, 0.23\,\% of a radiation length) with two cylindrical layers (at radii of 3.9 and 7.6 cm, 2.3\,\% of a radiation length) and covers the pseudorapidity ranges \etain{2} and \etain{1.4} for the inner and outer layers, respectively. The number of inactive (dead or noisy) individual pixels is small, about 1.5\,\%, but in addition some 17\,\% of the total area is currently not active, mostly because of insufficient cooling flow in some of the detector modules. 
The number of noise hits in the active pixels of the SPD was measured with a random trigger to be of the order of 
$10^{-4}$ per event.
The SPD was aligned using cosmic-ray tracks~\cite{alignement} collected prior to the collider run and tracks from collisions recorded at $\cms = \unit[0.9]{TeV}$.

Information from two scintillator hodoscopes, called VZERO counters, was used for event selection and background rejection. These counters are placed on either side of the interaction region at $z = 3.3$ m and $z = -0.9$ m. They cover the regions $2.8<\eta<5.1$ and $-3.7<\eta<-1.7$ and record both amplitude and time of signals produced by charged particles.

The central detector subsystems are placed inside a large solenoidal magnet which provides a field of $0.5$~T. For the 2.36 TeV data taking the VZERO detectors were not turned on. Therefore, the trigger conditions, the analysis and the systematic errors differ slightly between the two data sets (see below).

Because of the low interaction rate it was possible to use a rather loose trigger for collecting data. At 0.9 TeV, the minimum-bias trigger required a hit in either one of the VZERO counters or in the SPD detector; i.e. essentially at least one charged particle anywhere in the 8 units of pseudorapidity covered by these trigger detectors.
At the higher energy, the trigger required at least one hit in the SPD detector (\etain{2}). The events were collected in coincidence with the signals from two beam pick-up counters (BPTX), one on each side of the interaction region, indicating the presence of passing bunches.

The bunch intensity was typically $5 \times 10^{9}$ protons, giving a luminosity of the order of $\unit[10^{26}]{cm^{-2}s^{-1}}$. This value corresponds to a rate of a few Hz for inelastic proton--proton collisions  and a negligible pile-up probability for events in the same bunch crossing.

In the case of the $0.9$~TeV data, events in coincidence with only one passing bunch, as well as when no bunch was passing through the detector, were also registered. These control triggers were used to measure the beam-induced and accidental backgrounds.

The observed longitudinal sizes of the interaction regions can be inferred from Fig.~\ref{figVtx}, which shows the distribution of the interaction vertices along the beam axis reconstructed from hit correlation in the two SPD layers. The vertex distribution has an estimated r.m.s. of $\unit[4.1]{cm}$ and $\unit[2.7]{cm}$ for the 0.9~TeV and 2.36~TeV samples, respectively. These vertex distributions are for all triggered events with a reconstructed primary vertex after background rejection. They are compared to Monte Carlo simulations of proton--proton collisions using a Gaussian beam profile with a standard deviation as measured from data. The experimentally observed tails (mainly from \linebreak events with one or two reconstructed tracks) are well described in the simulation, confirming that beam induced background is very small in the selected sample. The ver\-tex-reconstruction efficiency is practically independent of the vertex $z$-position for $|z|<10$~cm.

\section{Event selection and corrections to INEL and NSD event classes}

Slightly different event selections were applied after data reconstruction for the analysis of the two collision energies because of the different detector configurations.

For both data samples, an offline selection is applied to reject beam induced background. At 0.9 TeV, the VZERO counters were used to remove beam--gas or beam-halo \linebreak events by requiring their timing signals, if present, to be compatible with particles produced in collision events (see \cite{ALICEfirst} for more details). At both energies, this background was also rejected by exploiting the correlation between the number of clusters of pixel hits and the number of tracklets (short track segments in the SPD, compatible with the event vertex, as described below). From the analysis of our control triggers, we found that background events typically have a large number of pixel hits compared with the number of tracklets pointing to the reconstructed vertex.

\begin{figure}[t!]
\centering
\includegraphics[width=\linewidth]{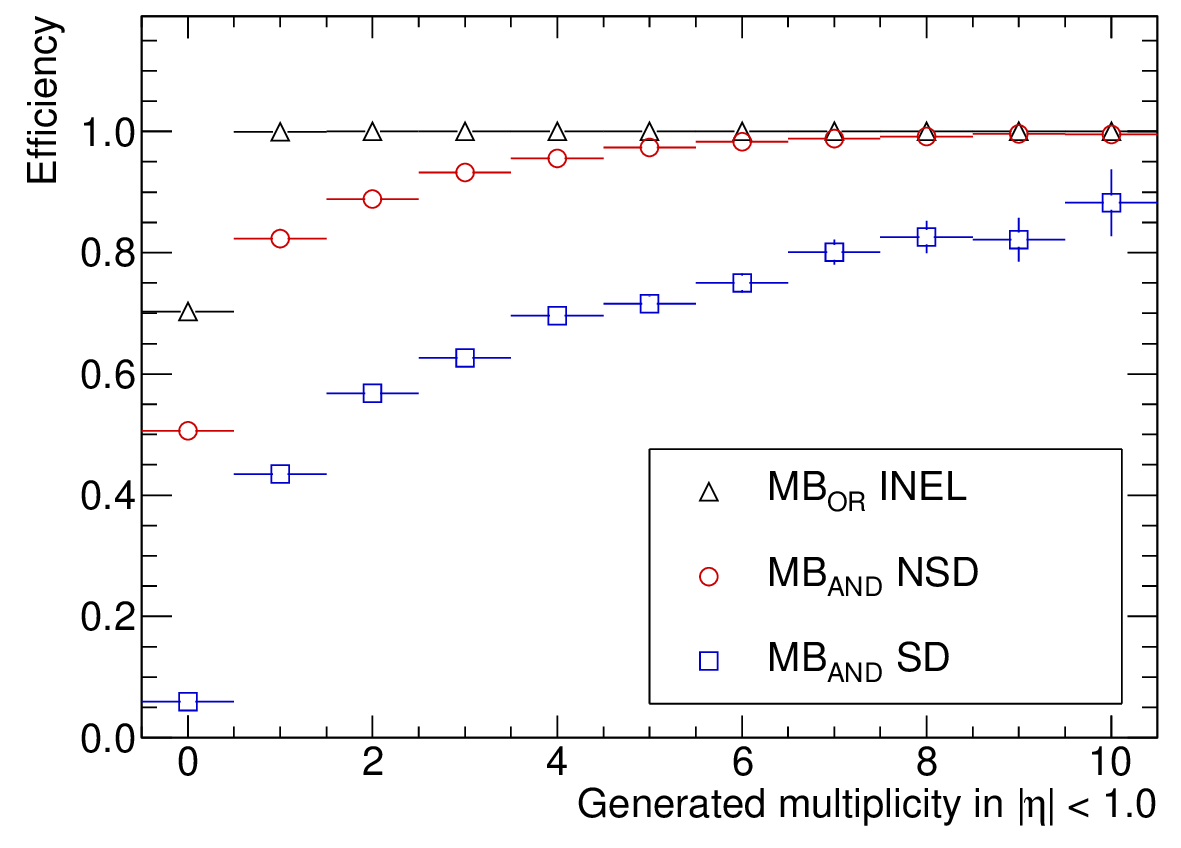}
\caption{Charged-particle multiplicity dependence of the selection efficiency, for INEL events (triangles) with MB$_{\mathrm{OR}}$ selection, and for NSD (circles) and SD (squares) events with MB$_{\mathrm{AND}}$ selection obtained with a simulation of the $0.9$~TeV data.}
\label{figverteff}
\end{figure}

\begin{table*}
\caption{(a) Relative fractions of SD and DD events, as obtained from previous measurements at 0.9 TeV \cite{UA5diff} and 1.8 TeV \cite{E710,CDF_DDD}. The measured DD fractions are scaled according to the prescription in~\cite{CDF_DDD}. Corresponding fractions calculated using PYTHIA and PHOJET are given for events within the diffractive-mass range covered experimentally (see text), and also without the restriction on diffractive-mass (parentheses). (b) Selection efficiencies for different classes of events: at 0.9~TeV, where the MB$_{\mathrm{OR}}$ selection was used for INEL sample and MB$_{\mathrm{AND}}$ for NSD sample; at 2.36 TeV, where the selection using the SPD only was used for both INEL and NSD samples.}
\label{tab:SD_vs_DD}
  \centering
  \begin{tabular}{lcccccc}
    \hline
    \tblspc
     & \multicolumn{6}{c}{(a) Relative process fractions}\\
    \hline
    \tblspc
     &\multicolumn{3}{c}{0.9 TeV}  &\multicolumn{1}{c}{1.8 TeV} & \multicolumn{2}{c}{2.36 TeV}  \\
     \cline{2-4}
     \cline{6-7}
    \tblspc
                            & Data \cite{UA5diff} & PYTHIA & PHOJET & Data \cite{E710,CDF_DDD}& PYTHIA & PHOJET \\
    \hline
    \tblspc
    SD  & $0.153 \pm 0.023$ & 0.189 (0.223) & 0.152 (0.191) & $0.159 \pm 0.024$ & 0.167 (0.209) & 0.126 (0.161)\\
    DD  & $0.095 \pm 0.060$ & 0.123 & 0.066 & $0.107 \pm 0.031$ & 0.127 & 0.057 \\
    \hline
    \tblspc
    &\multicolumn{6}{c}{(b) Selection efficiencies}\\
    \hline
    \tblspc
    &\multicolumn{4}{c}{0.9 TeV} &\multicolumn{2}{c}{2.36 TeV}  \\
    \cline{2-7}
    \tblspc
    & \multicolumn{2}{c}{PYTHIA} & \multicolumn{2}{c}{PHOJET} & PYTHIA & PHOJET\\
    \cline{2-5}
    \tblspc
    & MB$_{\mathrm{OR}}$ & MB$_{\mathrm{AND}}$ & MB$_{\mathrm{OR}}$ & MB$_{\mathrm{AND}}$ & MB$_{\mathrm{SPD}}$ & MB$_{\mathrm{SPD}}$ \\

    \hline
    \tblspc
    SD      & 0.77 & 0.29 & 0.86 & 0.34 & 0.55 & 0.62 \\
    DD      & 0.92 & 0.49 & 0.98 & 0.77 & 0.63 & 0.79 \\
    ND      & 1.00 & 0.98 & 1.00 & 0.96 & 0.99 & 0.99 \\
    INEL    & 0.95 &      & 0.97 &      & 0.86 & 0.90 \\
    NSD     &      & 0.92 &      & 0.94 & 0.94 & 0.97 \\
    \hline
  \end{tabular}
\end{table*}

At 0.9 TeV, for the INEL analysis, we used the triggered event sample requiring a logical OR between the signals from the SPD and  VZERO detectors (MB$_{\mathrm{OR}}$). However, for the NSD analysis we selected a subset of the total sample by requiring a coincidence between the two sides of the VZERO detectors (MB$_{\mathrm{AND}}$). This requires the detection of at least one charged particle in both the forward and backward hemispheres, which are separated by 4.5 units of pseudorapidity. In this subset, single-diffraction events are suppressed, therefore, model dependent corrections and associated systematic errors are reduced (see below). The selection efficiencies, MB$_{\mathrm{OR}}$ for INEL events and MB$_{\mathrm{AND}}$ for NSD events, are multiplicity dependent as illustrated in Fig.~\ref{figverteff}. As expected, the MB$_{\mathrm{AND}}$ selection has a low efficiency for SD events, in particular at low multiplicities, where they contribute most.
After these selections, the remaining background at 0.9 TeV was estimated, and corrected for, with the help of the control triggers. The background events (99\,\% of which have no reconstructed tracklets) correspond to about 2\,\% of the events in the INEL sample and to less than 0.01\,\% in the NSD sample.

The 2.36 TeV data sample was triggered by at least one hit in the SPD (MB$_{\mathrm{SPD}}$) and this selection was used for both INEL and NSD analyses. After rejecting the background using the correlation between the number of pixel hits and the number of tracklets, the remaining background (93\,\% of which has no reconstructed tracklets) was estimated to be 0.7\,\%. We have assumed that the correlation between the number of clusters of pixels and the number of tracklets is similar at both energies because accelerator and detector conditions did not change significantly between the two data collection periods.

In both data samples, the cosmic-ray contamination, estimated from the control triggers and from absolute rates, is negligible. Additional crosschecks of background levels were made by visual scanning of a few hundred selected events.

The number of collision events used in this analysis corresponds to about 150\,000 and 40\,000 interactions for the 0.9 and 2.36 TeV data, respectively.

The efficiencies of our selections and their sensitivities to variations in the relative fractions of event classes were studied using two different Monte Carlo generators, PYTHIA 6.4.14 and 6.4.21~\cite{Pythia,Pythia1} tune D6T \cite{D6Ttune} and PHOJET 1.12 \cite{PhoJet} used with PYTHIA 6.2.14, with the detector simulation and reconstruction software framework AliRoot~\cite{aliroot}, which includes a detailed model of the ALICE apparatus. Particle transport in the detector  was simulated using the \mbox{GEANT-3} software package~\cite{geant3}.

To normalize our results to INEL and NSD event class\-es, we used measured cross sections in p$\pbar$ collisions instead of those provided by the event generators. The single-diffraction production cross section at 0.9~TeV was measured by UA5 \cite{UA5diff} in the kinematic range $M^2 < 0.05s$, where $M$ is the mass of the diffractive system. The published value is a result of an extrapolation to low diffractive masses $M < 2.5$~GeV and therefore is model-dependent. There exist indications from measurements at lower energy, $\cms = 546$~GeV \cite{ua4_xsection}, and from phenomenological models \cite{Kaidalov:2009aw} that this cross section value may be underestimated by up to 30\,\%. We decided to use the published value, but we checked that our results stay within their systematic uncertainties, if instead we use a 30\,\% higher single-diffractive cross section.

At 1.8~TeV, the single-diffraction cross section was measured by E710 \cite{E710} and CDF \cite{CDF_SD}
in the kinematic ranges $\unit[2]{GeV^2} < M^2 < 0.05s$ and $\unit[1.4]{GeV^2} < M^2 < 0.15s$, respectively. As the CDF result includes significant model dependent acceptance corrections at low masses, we used the E710 measurement. The inelastic cross section at 1.8~TeV, needed for normalization, was taken from~\cite{CDF811}. At both energies, the fraction of SD events in the Monte Carlo generators was normalized to the data in the mass regions covered by the corresponding experiments.

To simulate the SD events we used the two Monte Carlo event generators and rely on them to tag these events. In order to classify an event as diffractive, the diffractive mass should satisfy a coherence condition ($M^2$ $< K s$, where $K = 0.05$--$0.15$), which effectively provides a high-mass cut-off. PHOJET uses this coherence condition with $K = 0.15$ and we checked that, if we would further restrict diffractive masses using $K = 0.05$, our multiplicity results will practically not change. PYTHIA generates, with a low probability, SD events with very high diffractive masses. Therefore, we checked the stability of our results by imposing the tight coherence condition ($K = 0.05$) to generated SD events. This decreased the average multiplicities for the NSD samples by less then 2\,\%, well within our systematic uncertainties.

Measurements of double-diffraction cross sections are available from UA5~\cite{UA5diff} at 0.9 TeV and CDF~\cite{CDF_DDD} at \linebreak 1.8~TeV. Experimentally, DD events are defined by requiring a minimum pseudorapidity gap (of about 3 units), where no charged particles are detected. When implementing these experimental cuts in the event generators, the results were widely fluctuating and inconsistent with the measurements, possibly because the occurrence of large rapidity gaps is very sensitive to the model assumptions and process parameterizations. Therefore, for classification of DD events we used the process type information provided by the generators but we adjusted the fractions to the measured values. The values used take into account an increase of the DD fractions due to the pseudorapidity-gap definition as described in~\cite{CDF_DDD}. Note that the correction arising from unmeasured DD events is small, both because the cross section for DD is small and because the event selection efficiency is large in our samples.

The relative fractions of SD and DD events, as measured in \cite{UA5diff, CDF_DDD, E710} are summarized in Table~\ref{tab:SD_vs_DD}, along with our calculated trigger and selection efficiencies. The relative fractions for SD and DD vary very slowly with energy, therefore, we used the measurements available at 1.8~TeV for the 2.36~TeV sample.

\section{Analysis method}

The analysis method is based on using hits in the two SPD layers to form short track segments, or tracklets. This method is similar to that used by the PHOBOS experiment with the first heavy-ion data from RHIC~\cite{PHOBOStracklets}.
We start with the reconstruction of the position of the interaction vertex by correlating hits in the two silicon-pixel layers. The vertex resolution achieved with this simple method depends on the track multiplicity, and is typically $0.1$--$\unit[0.3]{mm}$ in
the longitudinal ($z$) and $0.2$--$\unit[0.5]{mm}$ in the transverse direction. For events with only one SPD tracklet, the $z$-vertex position is determined by the point of closest approach to the mean beam axis.
A vertex was reconstructed for 83\,\% of events in the MB$_{\mathrm{OR}}$ selection and for 93\,\% of events in the MB$_{\mathrm{AND}}$ selection. At the higher energy, in the MB$_{\mathrm{SPD}}$ selection 93\,\% of events have a vertex reconstructed.
Events with vertices within $|z|<\unit[10]{cm}$ are used in this analysis.

Using the reconstructed vertex as the origin, we calculate the differences in azimuthal ($\Delta \varphi$, bending plane) and polar ($\Delta \theta$, non-bending direction) angles of two hits, one in the inner and one in the outer SPD layer.
Hit combinations, called tracklets, are selected by a cut on the sum of the squares of $\Delta \varphi$ and $\Delta \theta$, with a cut-off at \unit[80]{mrad} and \unit[25]{mrad}, respectively.
The cut imposed on the difference in azimuthal angles would reject charged particles with a transverse momentum below $\unit[30]{MeV/{\it c}}$; however, the effective transverse-momentum cut-off is determined by particle absorption in the material and is approximately $\unit[50]{MeV/{\it c}}$.
If more than one hit in a layer matches a hit
in the other layer, only the hit combination with the smallest angular
difference is used.

For the pseudorapidity-density measurement, all events with vertex in the range $|z| < 10$~cm are used. For multi\-plicity-distribution measurements, the whole $\eta$-inter\-val \linebreak considered has to be covered by the acceptance of the SPD, for every event. Therefore, only events from a limited $z$-range of collision vertices are used for the two largest $\eta$-intervals, which reduces the available event statistics. At 0.9~TeV these reductions are 15\,\% for \etain{1.0} and 60\,\% for \etain{1.3}, and at 2.36~TeV 4\,\% for \etain{1.0} and 46\,\% for \etain{1.3}.

The number of primary charged particles is estimated by counting
the number of tracklets. This number was corrected for:
\begin{itemize}
\item{geometrical acceptance, detector and reconstruction efficiencies;}
\item{contamination by weak-decay products of long-lived particles (K$^0_{\rm s}$, $\Lambda$, etc.), gamma conversions and  secondary interactions;}
\item{undetected particles below the $\unit[50]{MeV/{\it c}}$ transverse-momentum cut-off;}
\item{combinatorial background caused by an accidental association of hits in the two SPD layers, estimated from data by counting pairs of hits with a large $\Delta \varphi$.}
\end{itemize}

The probability of an additional collision in the same bunch crossing (pile-up) at the estimated luminosity is below $10^{-3}$.
The effect on both multiplicity density and multiplicity distribution measurements due to such events has been found to be negligible.
Particular attention was paid to events having zero or one charged tracklets in the SPD acceptance. For the 0.9~TeV sample, the number of zero-track events for $|z| < 10$~cm was estimated using triggered events without a reconstructed vertex.
At 2.36~TeV, due to the different trigger (see Section 2), we have to use Monte Carlo simulations to estimate this number and therefore the results are more model-dependent than those at 0.9~TeV. As a consequence, the size of systematic uncertainties on average multiplicity is bigger at 2.36~TeV than that at 0.9~TeV, as described in Section 5.

\bfig[b]
  \includegraphics[width=\columnwidth]{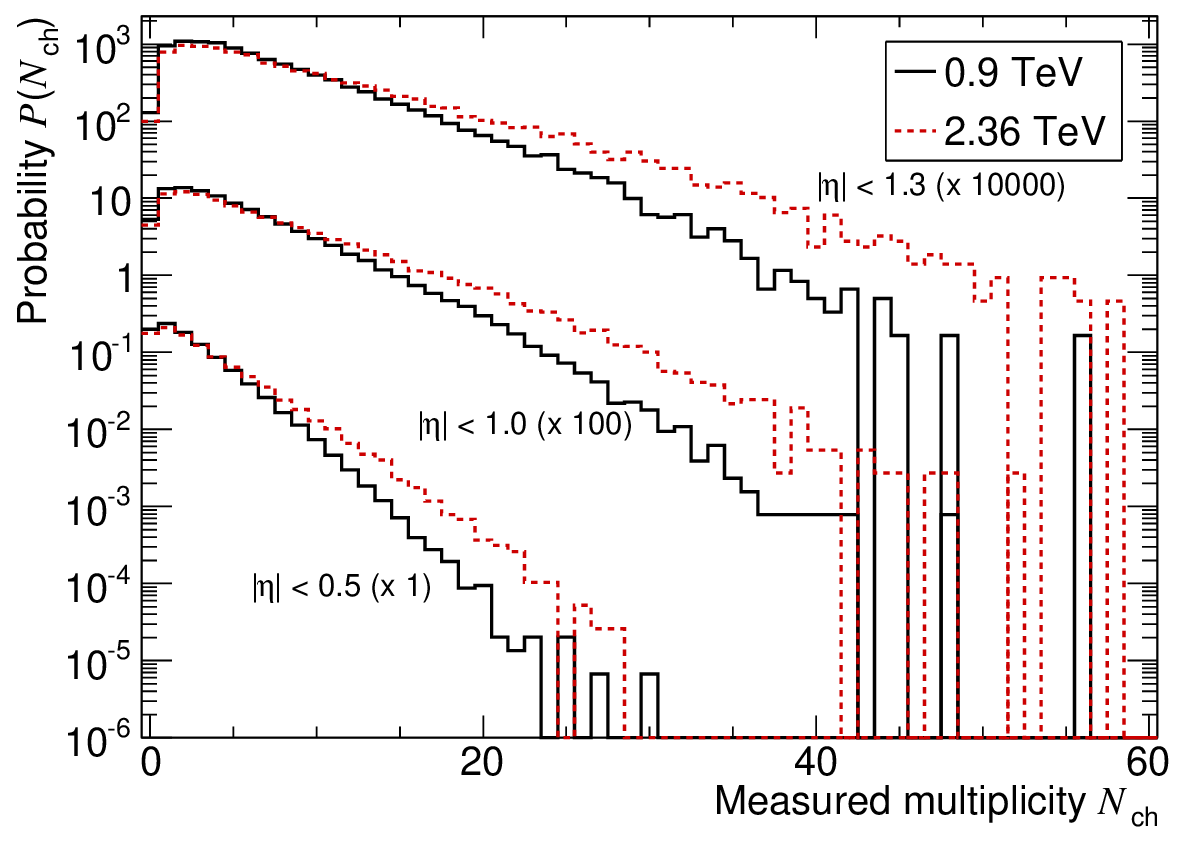}
  \caption{Measured raw multiplicity distributions in three pseudorapidity ranges for both energy samples ($0.9$~TeV full lines, $2.36$~TeV dashed lines). Note that for \etain{1.0} and \etain{1.3} the distributions have been scaled for clarity by the factor indicated.}
  \label{figure_measured}
\efig

The total number of collisions used for the normalization was 
calculated from the number of events with reconstructed vertex 
selected for the analysis and the number of triggered events without 
vertex. The latter number was corrected for beam-induced and 
accidental background as measured with the control triggers (see Section~2).
A small correction, determined from simulations, is applied to the number of tracks due to events with no reconstructed vertex.
In order to get the normalization for INEL and NSD events, we further corrected the number of events for the selection efficiency for these two event classes. For NSD events, we subtracted the single-diffrac\-ti\-ve contribution. The selection efficiencies depend on the char\-ged-particle multiplicity, as shown in
Fig.~\ref{figverteff} for the 0.9~TeV data sample for different event classes (INEL, NSD, and SD).
At both energies, the efficiency is close to 100\,\% for multiplicities of one or above for the INEL class, and reaches 90\,\% for multiplicities above two for the NSD class.
The averaged combined corrections in number of events due to the vertex-reconstruction and the selection efficiencies for INEL collisions are 5\,\% and 24\,\% for 0.9~TeV and 2.36~TeV data, respectively. This correction is larger at the higher energy because of significantly smaller pseudorapidity coverage of the MB$_{\mathrm{SPD}}$ selection compared with the MB$_{\mathrm{OR}}$ selection and the necessity for large correction for zero-multiplicity events at this energy. 
For NSD collisions, at both energies, these event-number corrections are small (2\,\% and 1\,\% for 0.9~TeV  and 2.36~TeV data, respectively) as a consequence of partial cancelation between adding non-observed ND and DD events, and subtracting triggered SD events. The resulting model-dependent correction factors due to the selection efficiencies applied to averaged charged-particle multiplicities for the NSD samples are 0.973 and 1.014 for 0.9~TeV and 2.36~TeV data, respectively.

\bfig[t]
  \includegraphics[width=\columnwidth]{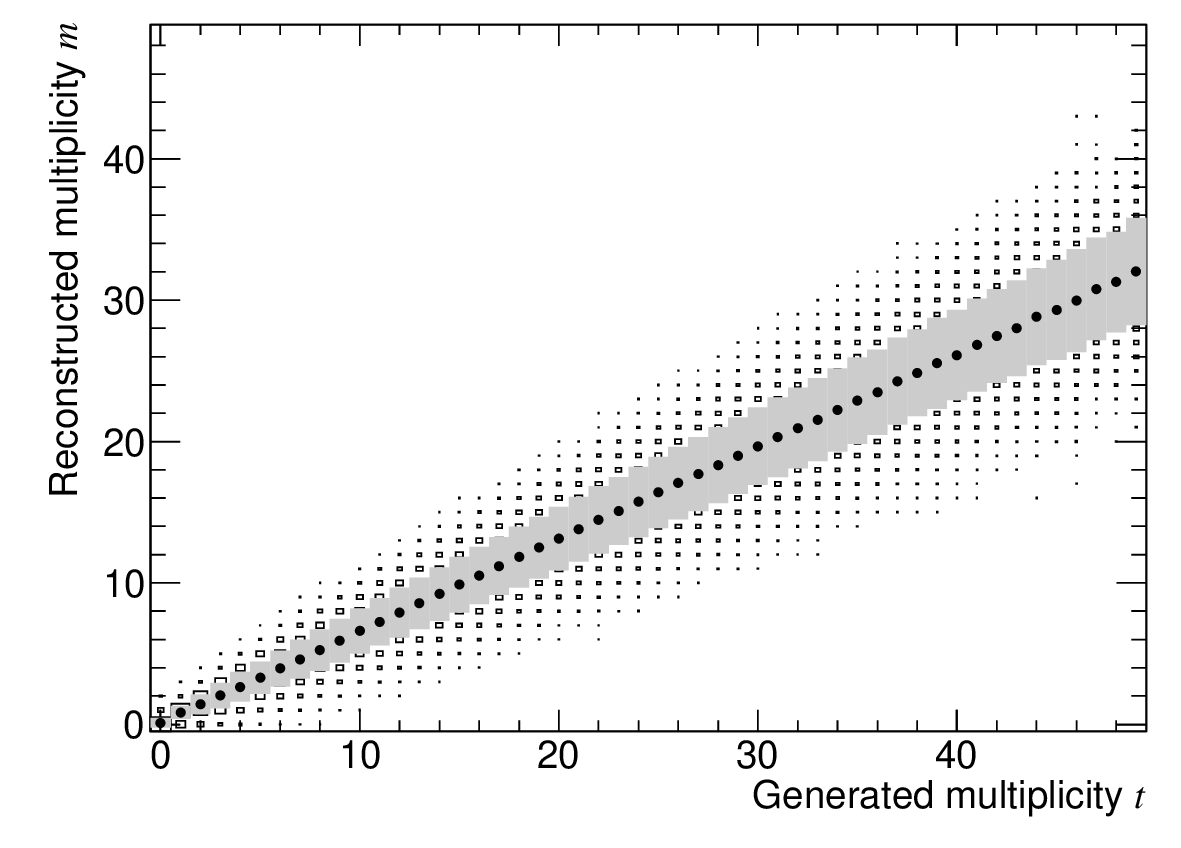}
  \caption{Graphical representation of the detector response matrix: number of tracklets found in the SPD ($m$) vs the number of generated
  primary particles in $|\eta| < 1.0$ ($t$) for $\cms = 0.9$~TeV.
  The distribution of the measured tracklet multiplicity for a given generated multiplicity shown with its most probable value (dots), r.m.s. (shaded areas), and full spread (squares).}
  \label{figure_responsematrix}
\efig

The multiplicity distributions, measured in three $\eta$-intervals, are shown in Fig.~\ref{figure_measured} for raw data at both energies.  The method used to correct the raw measured distributions for efficiency, acceptance, and other detector effects, is based on unfolding with $\chi^2$ minimization with regularization \cite{blobel_unfolding}. The detector response was determined with the same Monte Carlo simulation as described above. Figure~\ref{figure_responsematrix} illustrates the detector response matrix $R_{mt}$ for $|\eta| < 1$, which gives the conditional probability that a collision with multiplicity $t$ is measured as an event with multiplicity $m$. Therefore, each column is normalized to unity.
This matrix characterizes the properties of the detector and does not depend on the specific event generator used for its determination, apart from second-order effects due to, for example, differences in particle composition and momentum spectra, discussed in Section 5. As this matrix is practically independent of energy, it is shown for the $0.9$~TeV case only. The unfolded spectrum $U(N_{\rm ch})$ is found by minimizing
\bq
  \hat{\chi}^2(U) = \sum\limits_m \left(\frac{M_m - \sum_t R_{mt} U_t}{e_m} \right)^2  + \beta F(U) \label{chi2} ,
\eq
 where $R$ is the response matrix, $M$ is the measured spectrum, $e$ is the estimated measurement error, and $\beta F(U)$ is a regularization term that suppresses high-frequency components in the solution. The only assumption made about the shape of the corrected spectrum is that it is smooth. The smoothness is imposed by the choice
\begin{eqnarray}
  F(U) = \sum\limits_t \frac{(U'_t)^2}{U_t}
       = \sum\limits_t  \frac {(U_{t-1} - U_{t})^2} {U_t} ,
       \label{regularization}
\end{eqnarray}
which minimizes the fluctuations with respect to a constant constraint imposed by first derivatives.
The regularization coefficient $\beta$ is chosen such that, after minimization, the contribution of the first term in Eq.~\ref{chi2} is of the same order as the number of degrees of freedom (the number of bins in the unfolding).

The unfolded spectrum is corrected further for vertex reconstruction and event selection efficiencies (see Fig.~\ref{figverteff})

\bfig[t]
  \includegraphics[width=\columnwidth]{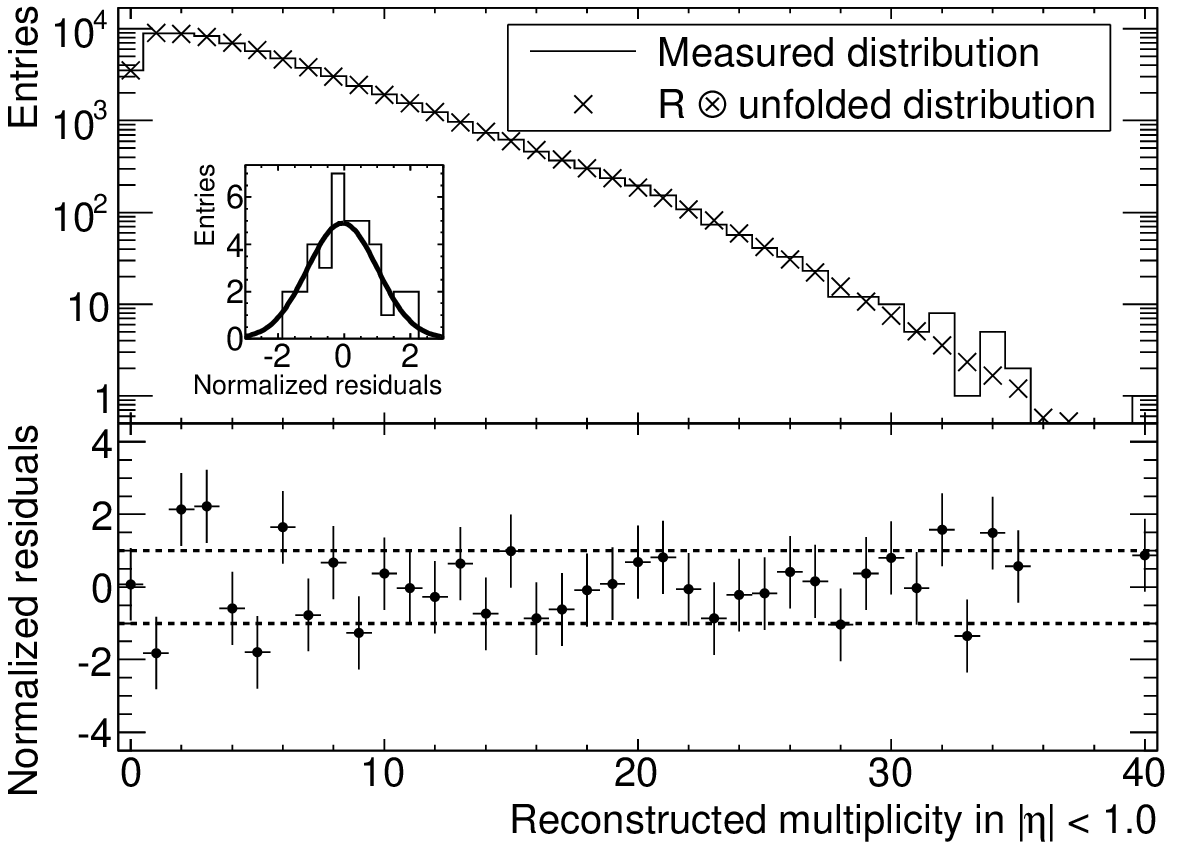}
  \caption{Measured raw multiplicity distribution
      (elements of vector $M$, histogram), superimposed on the convolution $RU$ of the unfolded distribution with the response matrix (crosses), at $\cms = 0.9$~TeV for $|\eta| < 1.0$ (upper plot). The error bars are omitted for visibility.
      Normalized residuals, i.e. the difference between the measured raw distribution and the corrected distribution folded with the response matrix divided  by the measurement error (lower plot).
       The inset shows the distribution of these normalized residuals fitted with a Gaussian ($\sigma \approx 1.06$).}
  \label{figure_residuals}
\efig

\begin{table*}[tb]
\centering
\caption{Contributions to systematic uncertainties in the measurements of the charged-particle pseudorapidity density and of the multiplicity distribution. For pseudorapidity densities, when two values are given, they correspond to the pseudorapidities $0.0$ and $1.4$, respectively. The sign of the event-generator uncertainties indicates if the result using PHOJET corrections is higher (positive sign) or lower (negative sign) than that using PYTHIA corrections. For multiplicity distributions the values are given for \etain{1.0}. Multiple values indicate uncertainties for respective multiplicities shown in parentheses.
}
\label{systable}
\begin{tabular}{lllll}
	\hline
    \tblspc
	Uncertainty					&  \multicolumn{2}{l}{$\dNdEta$ analysis} &   \multicolumn{2}{l}{$P(N_\mathrm{ch})$ analysis} \\
	\cline{2-5}
    \tblspc
    & $0.9$~TeV & $2.36$~TeV & $0.9$~TeV & $2.36$~TeV \\
    \hline
    \tblspc
    Tracklet selection cuts                 & negl.     & negl.     & negl.     & negl. \\
    Material budget                         & negl.     & negl.     & negl.     & negl. \\
    Misalignment                            & negl.     & negl.     & negl.     & negl. \\
    Particle composition                    & 0.5--1.0\,\%	    & 0.5--1.0\,\%     & \multicolumn{2}{c}{included in detector efficiency}	 \\
    Transverse-momentum spectrum            & 0.5\,\%   & 0.5\,\%   & \multicolumn{2}{c}{included in detector efficiency} \\
	Contribution of diffraction (INEL)        & 0.7\,\%  & 2.6\,\% &  3--0\,\% (0--5)	   & 5--0\,\% (0--5)	\\
	Contribution of diffraction (NSD)       & 2.8\,\%  & 2.1\,\%   & 24--0\,\% (0--10)  & 12--0\,\% (0--10)\\
	Event-generator dependence (INEL)	    & $+1.7$\,\%     & $+5.9$\,\%    & 8--0\,\% (0--5)	    & 25--0\,\% (0--10)	\\
	Event-generator dependence (NSD)	    & $-0.5$\,\%     & $+2.6$\,\%    & 3--5--1\,\%  (0--10--40)     & 32--8--2\,\% (0--10--40)\\
    Detector efficiency                     & 1.5\,\%   & 1.5\,\% & 2--4--15\,\% (0--20--40) & 3--0--9\,\% (0--8--40) \\
    SPD triggering efficiency               & negl.     & negl.    & negl.  & negl. \\
    VZERO triggering efficiency (INEL)        & negl.  & n/a  & negl. & n/a \\
    VZERO triggering efficiency (NSD)         & 0.5\,\%& n/a & 1\,\% & n/a \\
	Background events    		            & negl. & negl. & negl. & negl. \\
	\hline
    \tblspc
	Total (INEL)			                & $^{+2.5}_{-1.8}$\,\% & $^{+6.7}_{-3.1}$\,\% & 9--4--15\,\% (0--20--40) & 25--0--9\,\% (0--10--40) \\
	Total (NSD) 			                & $^{+3.3}_{-3.3}$\,\% & $^{+3.7}_{-2.7}$\,\% & 24--5--15\,\% (0--10--40) & 32--8--9\,\% (0--10--40) \\
	\hline
\end{tabular}
\end{table*}

The behaviour of the deconvolution method is illustrated in
 Fig.~\ref{figure_residuals} for the case \etain{1.0} at $\cms = 0.9$~TeV showing that the normalized residuals are well-behaved over the whole measured multiplicity range. The $\chi^2$ difference between the measured raw distribution and the corrected distribution folded with the response matrix is $\chi^2/ndf = 36.7/35 = 1.05$. Similar behaviour is observed for other $\eta$ intervals and at 2.36~TeV.

We checked the sensitivity of our results to --
\begin{itemize}
\item The value of the regularization coefficient $\beta$.
\item Changing the regularization term, defined in Eq.~\ref{regularization}, to:
\begin{eqnarray}
  F(U) = \sum\limits_t \frac{(U''_t)^2}{U_t}
       = \sum\limits_t  \frac {(U_{t-1} - 2 U_{t} + U_{t+1})^2} {U_t} ,
       \label{regularization2}
\end{eqnarray}
which minimizes the fluctuations with respect to a linear constraint imposed by second derivatives.
\item Changing the unfolding procedure. An unfolding based on Bayes' theorem \cite{agostini_bayes, agostini_yellowreport} produces consistent results. It is an iterative procedure using the relations:
\bq
  \tilde{R}_{tm} = \frac{ R_{mt} \cdot P_t }{ \sum_{t'} R_{mt'} P_{t'} } , \hspace{1cm} U_t = \sum\limits_{m} \tilde{R}_{tm} M_m , \label{eq_mult_bayesian}
\eq
with an \emph{a priori} distribution $P$. The result $U$ of an iteration is used as a new \emph{a priori} $P$ distributions for the following iteration.
\item Variation of convergence criteria and initial distribution. For both unfolding procedures we checked that the results are insensitive to the details of the convergence criteria and a reasonable choice of initial distributions.
\end{itemize}

The details of this analysis are described in~\cite{JanFiete}.

\section{Systematic uncertainties}

In order to estimate the systematic uncertainties, the abo\-ve analysis was repeated:
\begin{itemize}
\item{varying the $\Delta\varphi$ and $\Delta\theta$ cuts used for the tracklet definition by $\pm 20$\,\%;}
\item{varying the density of the material in the tracking system, thus changing the material budget by $\pm 10$\,\%;}
\item{allowing for detector misalignment by an amount of up to $100 \, \mu$m;}
\item{varying the composition of the produced particle types with respect to the yields suggested by the event generators by $\pm 30$\,\%;}
\item{varying the non-observed-particle yield below the transverse momentum cut-off for tracklet reconstruction by $\pm 30$\,\%;}
\item{varying  the ratios of the ND, SD, and DD cross sections according to their measured values and errors shown in Table~\ref{tab:SD_vs_DD}, thus evaluating the uncertainty in the normalization to INEL and NSD events;}
\item{varying the thresholds applied to VZERO counters, both in simulation and in data (for the 0.9~TeV sample).}
\end{itemize}
The results are summarized in Table~\ref{systable} using the corrections calculated with PYTHIA tune D6T. Whenever corrections obtained with PHOJET give a different value, the difference is used in calculating an asymmetric systematic uncertainty. These two models were chosen because they predict respectively the lowest and the highest charged-particle densities for INEL collisions at both energies (see Section 6).

\begin{figure*}[htb]
\centering
\includegraphics[width=0.49\linewidth]{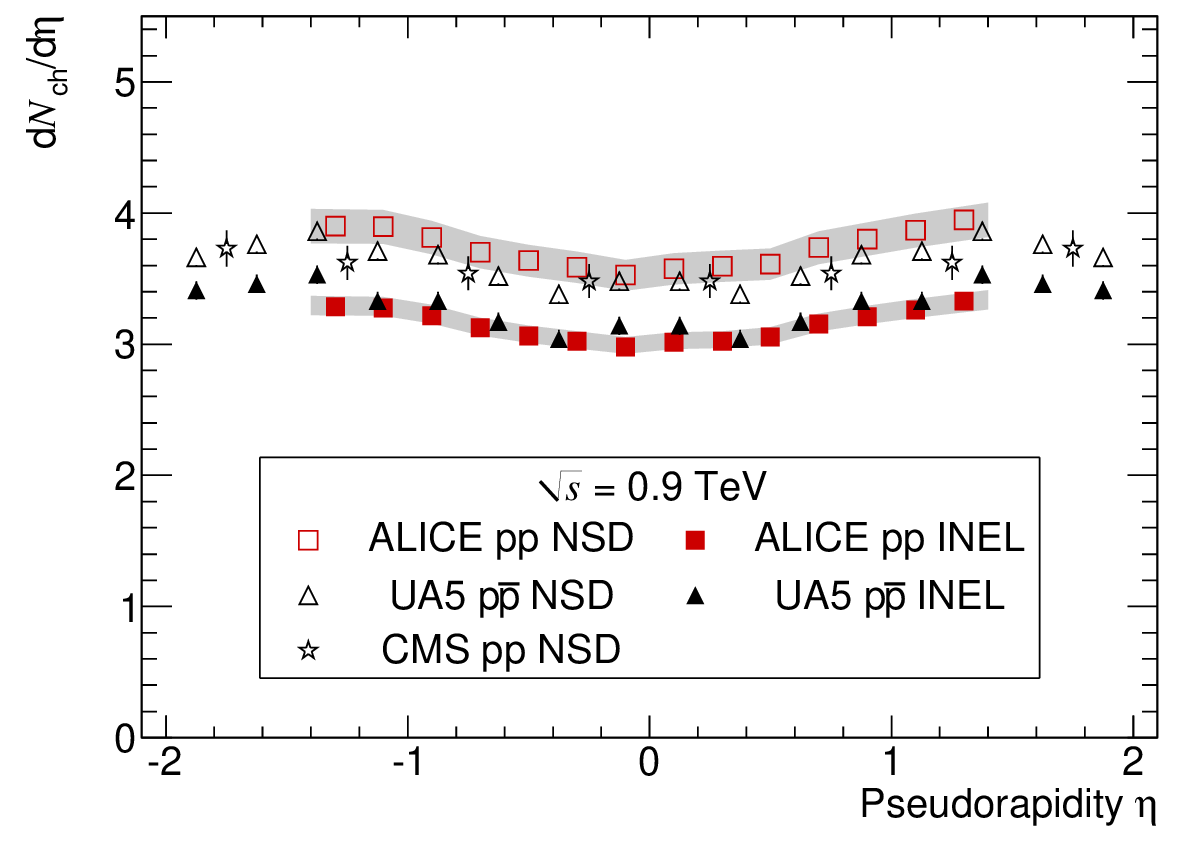}
\includegraphics[width=0.49\linewidth]{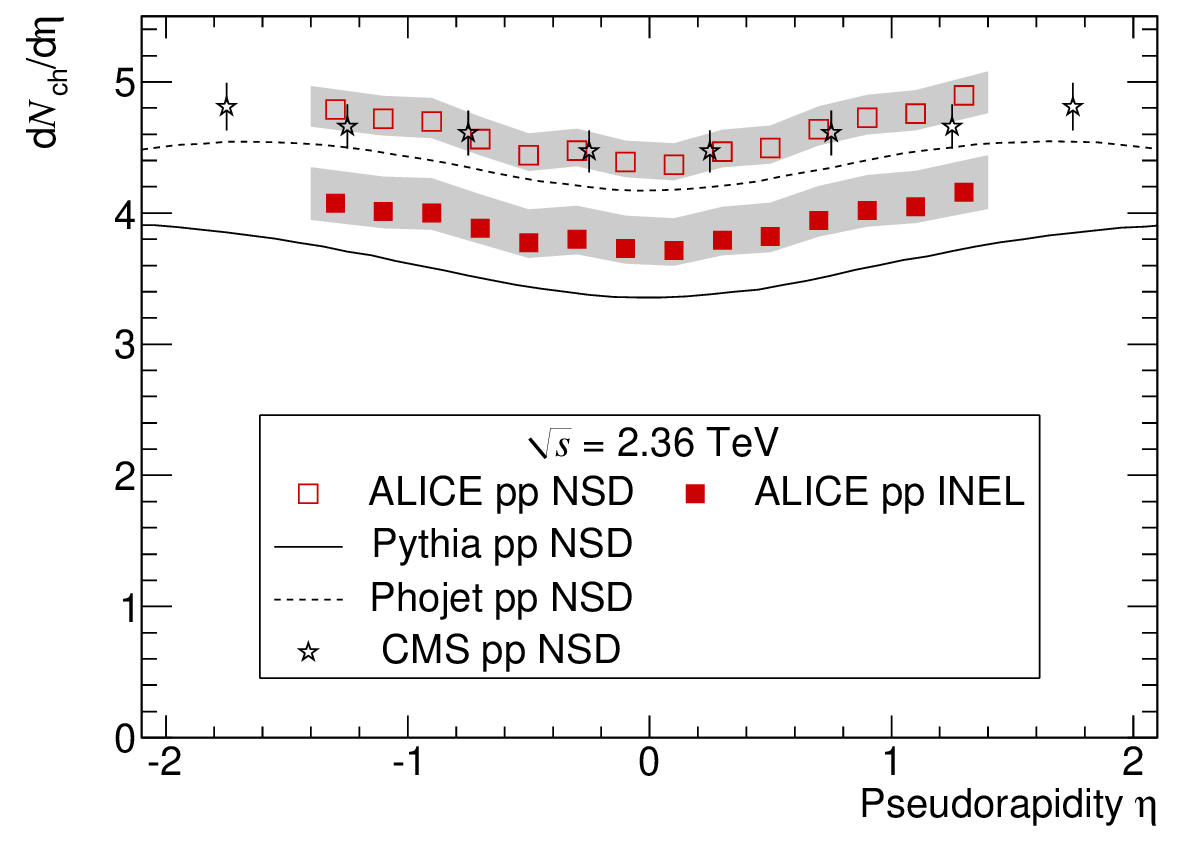}
\caption{
  Left: measured pseudorapidity dependence of $\dNdEta$ at $\cms = 0.9$~TeV for INEL (full symbols) and NSD (open symbols) collisions. The ALICE measurements (squares) are compared to UA5 p$\pbar$ data~\cite{UA5} (triangles) and to CMS pp data at the LHC \cite{CMS_first} (stars).
 Right: measured pseudorapidity dependence of $\dNdEta$ at $\cms = 2.36$~TeV for INEL (full symbols) and NSD (open symbols) collisions. The ALICE measurement (squares) for NSD collisions is compared to CMS NSD data \cite{CMS_first} (stars) and to model predictions, PYTHIA tune D6T~\cite{D6Ttune} (solid line) and PHOJET \cite{PhoJet} (dashed line). For the ALICE data, systematic uncertainties are shown as shaded areas; statistical uncertainties are invisible (smaller than data marks). For CMS data error bars show the statistical and systematic uncertainties added in quadrature.
 }
\label{figdNdEta}
\end{figure*}

The SPD efficiencies for trigger and for pixel hits are determined from the data. The SPD trigger efficiency is determined to be 98\,\% with negligible uncertainty based on analysis of the trigger information recorded in the data stream for events with more than one tracklet. 
The detector efficiency is determined from pixel-hit distributions, and checked by tracklet reconstruction. The uncertainty on the detector acceptance and efficiency due to the limited hit statistics and the current alignment precision of the detector is estimated by this method to be 1.5\,\%. 
The uncertainty in background corrections was estimated according to the description in Section~3.

The total systematic
uncertainty on the pseudorapidity density measurement at 0.9~TeV is smaller than $2.5$\,\% for INEL collisions and is about $3.3$\,\% for NSD collisions. At 2.36~TeV, the corresponding uncertainties are below 6.7\,\% and 3.7\,\% for INEL and NSD collisions, respectively. For all cases, they are dominated by uncertainties in the cross sections of diffractive processes and their kinematics.

To evaluate the systematic error on the multiplicity distribution, a new response matrix was generated for each change listed above and used to unfold the measured spectrum. The difference between these unfolded spectra and the unfolded spectrum produced with the unaltered response matrix determines the systematic uncertainty.

Additional systematic uncertainties originate from the unfolding method itself, consisting of two contributions.
The first one arises from statistical fluctuations due to the finite number of events used to produce the response matrix as well as the limited number of events in the measurement. The unfolding procedure was repeated 100 times while randomizing the input measurement and the response matrix according to their respective statistical uncertainties.
The resulting uncertainty due to the response matrix fluctuations is negligible. The uncertainty on the measured multiplicity distribution due to the event statistics reproduces the uncertainty obtained with the minimization procedure, as expected.

A second contribution arises from the influence of the regularization on the distribution.
The bias introduced by the regularization was estimated using the prescription described in~\cite{regbias} and is significantly lower than the statistical error inferred from the $\chi^2$ minimization, except in the low-multiplicity region. In that region, the bias is about 2\,\%, but the statistical uncertainty is negligible. Therefore, we added the estimated value of the bias to the statistical uncertainty in this region.
The correction procedure is insensitive to the shape of the multiplicity distribution of the events, which produce the response matrix.

Table~\ref{systable} summarizes the systematic uncertainties for the multiplicity distribution measurements. Note that the uncertainty is a function of the multiplicity which is reflected by the ranges of values. Further details about the analysis, corrections, and the evaluation of the systematic uncertainties are in~\cite{JanFiete}.

Both the pseudorapidity density and multiplicity distribution measurements have been cross-checked by a second analysis employing the Time-Projection Chamber \linebreak
(TPC)~\cite{ALICEdet}. It uses tracks and vertices reconstructed in the TPC in the pseudorapidity region $|\eta| < 0.8$. The pseudorapidity density is corrected using a method similar to that used for the SPD analysis. The results of the two independent analyses are consistent.

\begin{table*}[t]
\centering
\caption{Charged-particle pseudorapidity densities measured by ALICE in the central pseudorapidity region (\etain{0.5}), for inelastic (INEL), non-single-diffractive (NSD), and inelastic with $N_{\mathrm{ch}} > 0 $ (INEL$>$0) proton--proton collisions at centre-of-mass energies of $\unit[0.9]{TeV}$ and $\unit[2.36]{TeV}$. The ratios of multiplicity densities between the two energies are also given. Data at $\cms=\unit[0.9]{TeV}$ are compared to CMS NSD data \cite{CMS_first} and UA5 NSD and INEL p$\pbar$ data \cite{UA5}. Data at $\cms=\unit[2.36]{TeV}$ are compared to CMS NSD data. For ALICE and CMS measurements, the first error is statistical and the second one is systematic; no systematic uncertainty is quoted by UA5. These data are also compared to predictions for ${\rm p}{\rm p}$ collisions from different models: QGSM \cite{QGSMcal}, PYTHIA tune D6T \cite{D6Ttune} (\textit{a}), tune ATLAS-CSC \cite{CSCtune} (\textit{b}), and tune Perugia-0 \cite{Perugiatune} (\textit{c}), and PHOJET \cite{PhoJet}.}
\label{tabdNdeta}
\begin{tabular}{lllllllll}
  \hline
  \tblspc
  Experiment & ALICE pp & CMS pp & UA5 p$\pbar$ & QGSM & \multicolumn{3}{c}{PYTHIA} & PHOJET \\
  \cline{6-8}
  \tblspc
  Model   &  &  &  &  &  \textit{a} &   \textit{b} &   \textit{c} &  \\
  \hline
  \multicolumn{8}{c}{\tblspc $\cms = 0.9$~TeV} \\
  \hline
  \tblspc
  INEL &   $3.02 \pm 0.01 ^{+0.08}_{-0.05}$  &                          & $3.09 \pm 0.05$ &2.98&2.35&3.04&2.46&3.21\\
  NSD  &   $3.58 \pm 0.01 ^{+0.12}_{-0.12}$ & $3.48 \pm 0.02 \pm 0.13$ & $3.43 \pm 0.05$ &3.47&2.85&3.74&3.02&3.67\\
  INEL$>$0 &   $4.20 \pm 0.01 \pm 0.03$ &  &  &  & 3.40 & 4.35 & 3.61 & 4.06 \\
  \hline

  \multicolumn{8}{c}{\tblspc $\cms = 2.36$~TeV} \\
  \hline
  \tblspc
  INEL & $3.77 \pm 0.01 ^{+0.25}_{-0.12}$ &                           &   & 3.65&2.81&3.64&2.94&3.76\\
  NSD  & $4.43 \pm 0.01 ^{+0.17}_{-0.12}$ & $4.47 \pm 0.04 \pm 0.16$  &   & 4.14&3.38&4.44&3.57&4.20\\
  INEL$>$0 &   $5.13 \pm 0.03 \pm 0.03$ &  &  &  & 3.95 & 5.05 & 4.18 & 4.62 \\
  \hline

  \multicolumn{8}{c}{\tblspc Ratios} \\
  \hline
  \tblspc
  INEL & $1.247 \pm 0.005 ^{+0.057}_{-0.028}$ &                            &  & 1.22 & 1.20 & 1.20 & 1.20 & 1.17\\
  NSD  & $1.237 \pm 0.005 ^{+0.046}_{-0.011}$ & $1.28 \pm 0.014 \pm 0.026$ &  & 1.19 & 1.19 & 1.19 & 1.18 & 1.14\\
  INEL$>$0 &   $1.226 \pm 0.007 \pm 0.010$ &  &  &  & 1.16 & 1.16 & 1.16 & 1.14 \\
  \hline
\end{tabular}
\end{table*}

\section{Results}

In this section, pseudorapidity density and multiplicity distribution results are presented for two centre-of-mass energies and compared to results of other experiments and to models. For the model comparisons we have used QGSM~\cite{QGSMcal}, three different tunes of PYTHIA,
tune D6T~\cite{D6Ttune}, tune ATLAS-CSC~\cite{CSCtune} and tune Perugia-0~\cite{Perugiatune}, and PHOJET~\cite{PhoJet}. The PYTHIA tunes have been developed by three independent groups extensively comparing Monte Carlo distributions to underlying-event and minimum-bias Tevatron data.  Data from hadron colliders at lower energies have been used to fix the energy scaling of the parameters. Tune D6T uses the old PYTHIA multiple scattering and $Q^2$-ordered showers, whereas the two other tunes use the new multiple-scattering model provided by PYTHIA 6.4 and transverse-momentum-ordered showering.
Perugia-0 was not tuned for diffractive processes, \linebreak which affects the validity of this tune for the lowest multiplicities.
For final-state-radiation and hadronization, Peru\-gia-0 adds parameters fitted to LEP data.
The charged-particle density in the central rapidity region is mainly influenced by the infrared cut-off for parton scattering at the reference energy
(1.8~TeV) and its energy dependence.

Figure~\ref{figdNdEta} (left) shows the charged-particle density as a function of pseudorapidity obtained for INEL and NSD interactions at a centre-of-mass energy $\cms = 0.9$~TeV compared to p$\pbar$ data from the UA5 experiment~\cite{UA5}, and to pp NSD data from the CMS experiment~\cite{CMS_first}. The result is consistent with our previous measurement~\cite{ALICEfirst} and with UA5 and CMS data. Figure~\ref{figdNdEta} (right) shows the measurement of $\mathrm{d}N_\mathrm{ch}/\mathrm{d}\eta$  for INEL and NSD interactions at $\cms = 2.36$~TeV compared to CMS NSD data~\cite{CMS_first} and to PYTHIA tune D6T and PHOJET calculations. Our results for NSD collisions are consistent with CMS measurements, systematically above the PHOJET curve, and significantly higher than the distribution obtained with the PYTHIA tune D6T. Note that in the CMS pseudorapidity-density measurement the contribution from charged leptons was excluded. This implies that the CMS value is expected to be approximately 1.5\,\% lower than in our result, where charged leptons are counted as primary particles.

The pseudorapidity density measurements in the central region ($|\eta| < 0.5$) are summarized in Table~\ref{tabdNdeta} along with model predictions obtained with QGSM, PHOJET and three different PYTHIA tunes. Note that QGSM is not readily available as an event generator and the predictions for some of the event classes were obtained analytically by the authors of~\cite{QGSMcal}. At both energies, PYTHIA tune D6T and PHOJET yield respectively the lowest and highest charged-particle densities for INEL collisions.

\begin{figure}[b]
\centering \includegraphics[width=\linewidth]{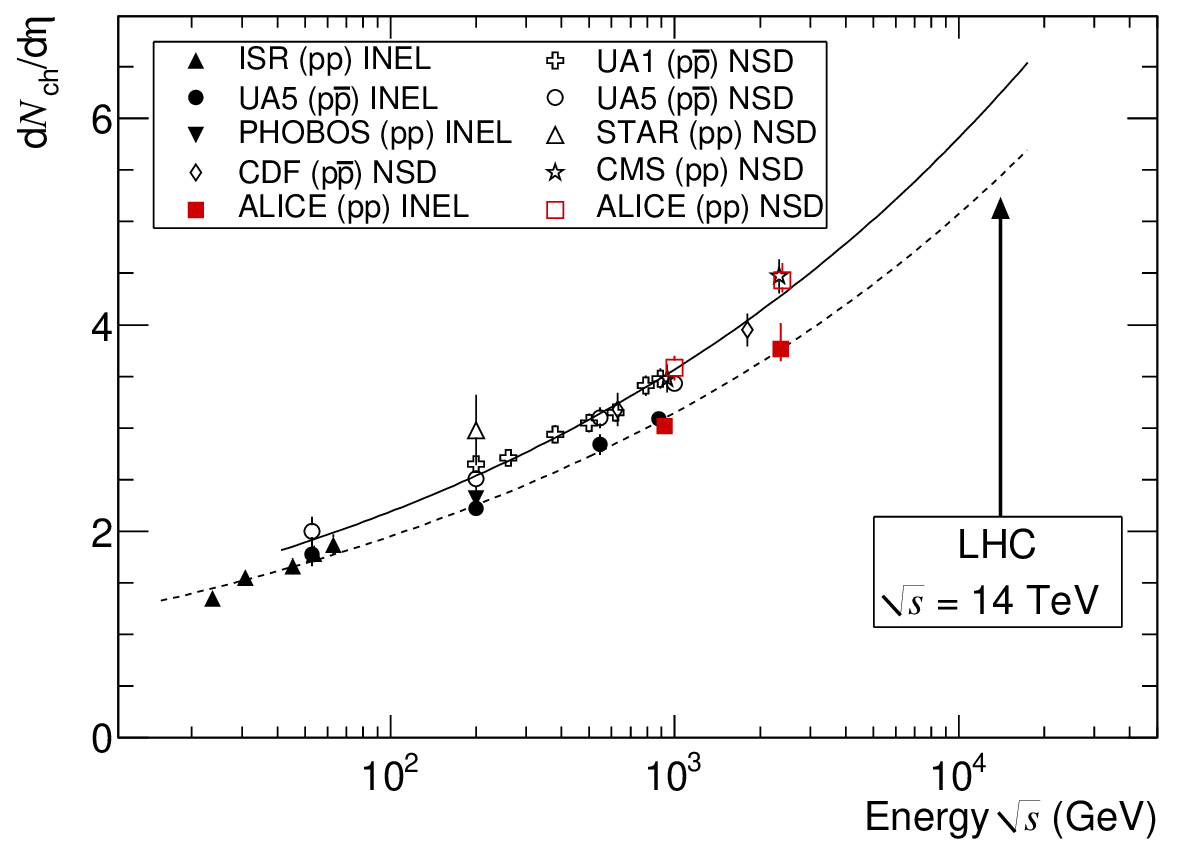}
\caption{Charged-particle pseudorapidity density in the central rapidity region
in proton--proton and proton--antiproton interactions as a function of
the centre-of-mass energy. The dashed and solid lines (for INEL and NSD interactions, respectively) show a
fit with a power-law dependence on energy. Note that data points at the same energy have been slightly shifted horizontally for visibility.}
\label{figdNdEtaSqrtS}
\end{figure}

\begin{figure*}[bt]
\centering
  \includegraphics[width=\columnwidth]{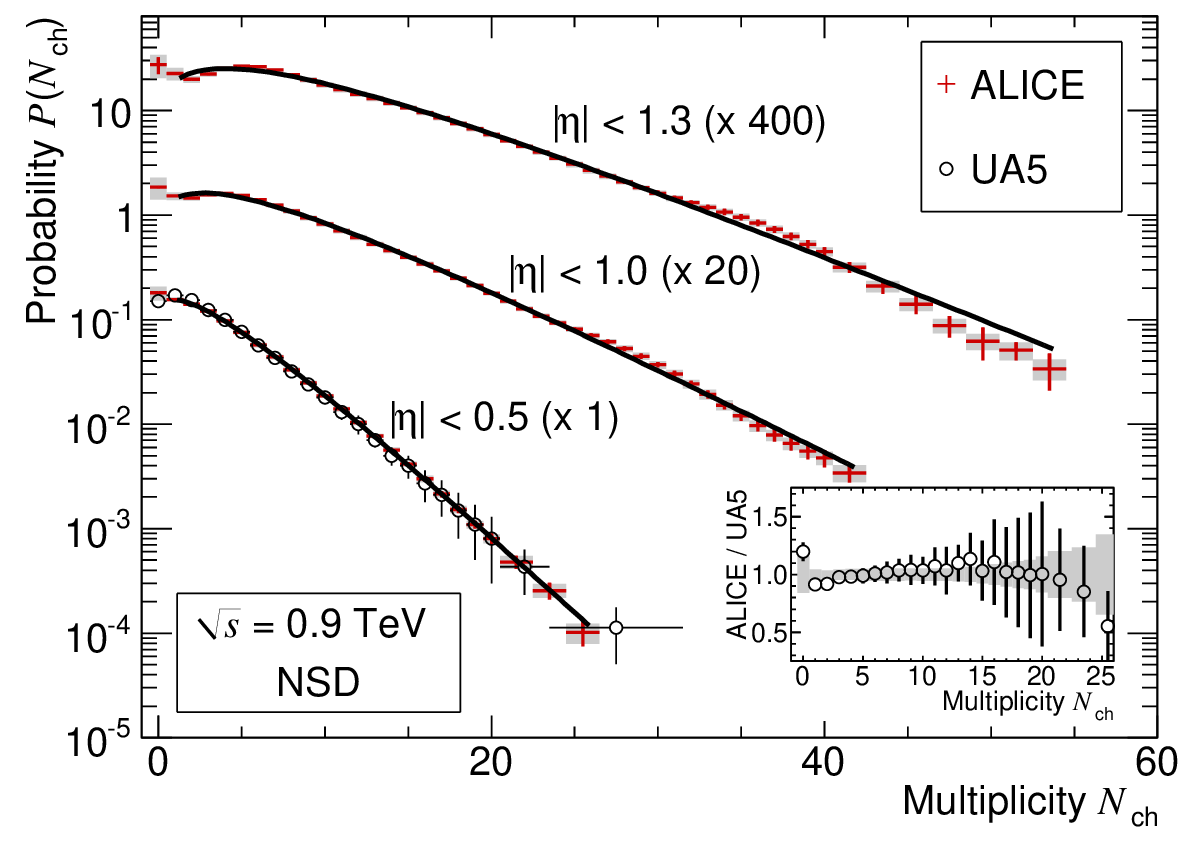}
  \includegraphics[width=\columnwidth]{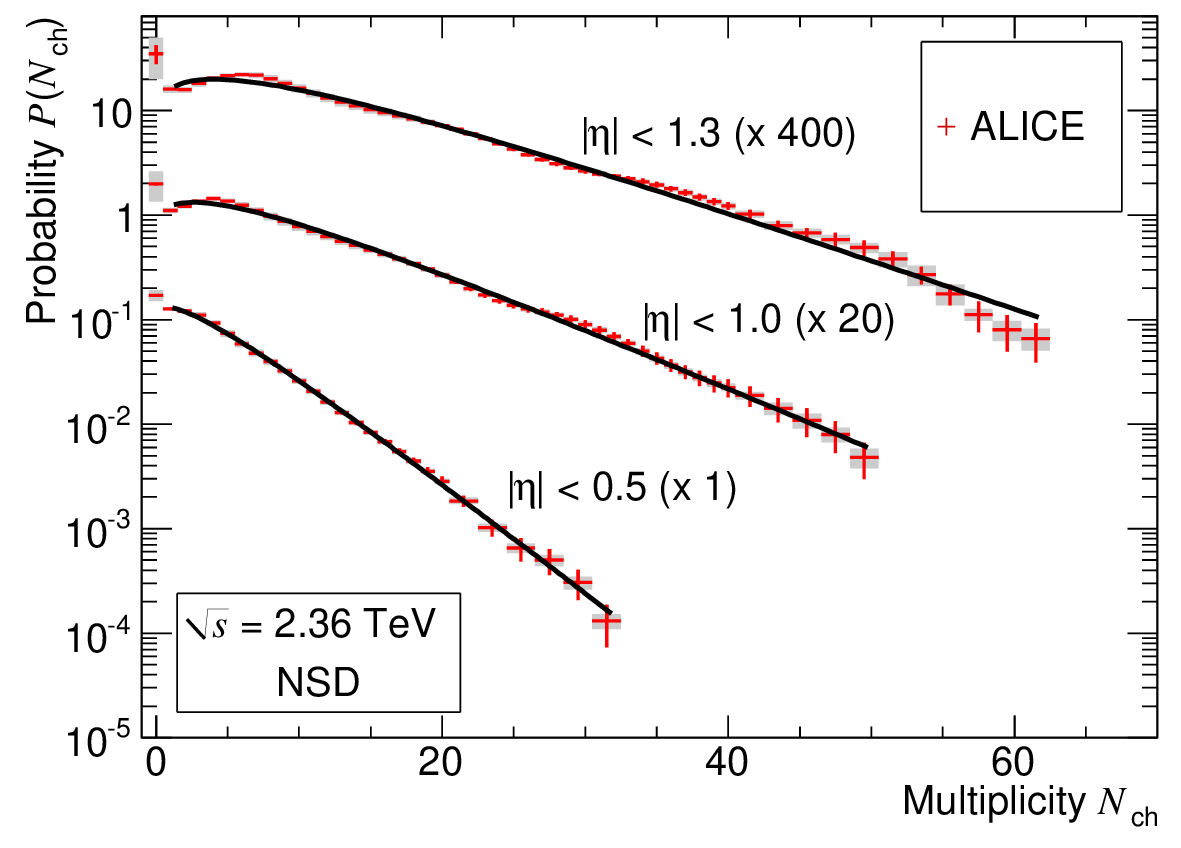}
  \caption{Corrected multiplicity distributions in three pseudorapidity ranges for NSD events. The solid lines show NBD fits. Error bars represent  statistical uncertainties and shaded area systematic ones. Left: data at $\cms = 0.9$~TeV. The ALICE measurement for \etain{0.5} is compared to the UA5 data at the same energy \cite{ua5_mult3}. In the inset the ratio of these two measurements is shown, the shaded area represents our combined statistical and systematic uncertainty, and the error bars those of UA5. Right: data at $\cms = 2.36$~TeV.
  Note that for \etain{1.0} and \etain{1.3} the distributions have been scaled for clarity by the factor indicated.}
  \label{figure_ua5}
\end{figure*}

Because part of the systematic uncertainties cancels in the ratio of the multiplicity densities between the two energies, these ratios are compared to model calculations as well. The main contribution to the systematic uncertainties in the measurement of charged-particle densities comes from the estimate of the number of events with zero tracks. Therefore, in addition to the two event classes (INEL and NSD) introduced so far, results are also presented for inelastic events with at least one charged particle produced in the region $|\eta|<0.5$, labeled as INEL$>$0. These values were obtained as the mean values of the corresponding corrected multiplicity distributions for $N_{\mathrm{ch}} > 0$ (see Fig.~\ref{figure_ua5}).

The consistency between data and model calculations varies with event class and the collision energy.
PYTHIA tunes D6T and Perugia-0 significantly underestimate the charged-particle density in all event classes and at both energies. ATLAS-CSC tune, PHOJET, and QGSM are closer to the data and describe the average multiplicity reasonably well, at least for some of the classes and energies listed in the Table~\ref{tabdNdeta}. However, the relative increase
in charged-particle density is underestimated by all models and tunes, most significantly for the event class with at least one charged particle in the central region (INEL$>$0). The increase predicted by all PYTHIA tunes is 16\,\% (14\,\% for PHOJET), whereas the observed increase is substantially larger ($22.6 \pm 0.7 \pm 1.0$)\,\%.

Figure~\ref{figdNdEtaSqrtS} shows the centre-of-mass energy dependence of the
pseudorapidity density in the central region.
The data points are obtained in the $|\eta|<0.5$ range from this experiment and from~\cite{UA5, ITS_dNdEta, R210,R210p, RHICRef, RHICRef1,UA5Rep, UA1, CDF_dNdEta}. When necessary, corrections were applied for differences in pseudorapidity ranges, fitting the pseudorapidity distributions around $\eta = 0$.

Using parameterizations obtained by fitting a power-law dependence on the centre-of-mass energy, extrapolations to the centre-of-mass energy of $\cms=\unit[7]{TeV}$ give \dndeta{4.7} and \dndeta{5.4} for INEL and for NSD
interactions, respectively. At
the nominal LHC energy of $\cms=\unit[14]{TeV}$, the same extrapolations yields
\dndeta{5.4} and \dndeta{6.2} for INEL and for NSD
collisions, respectively.

\begin{figure*}[tb]
\centering
  \includegraphics[width=\columnwidth]{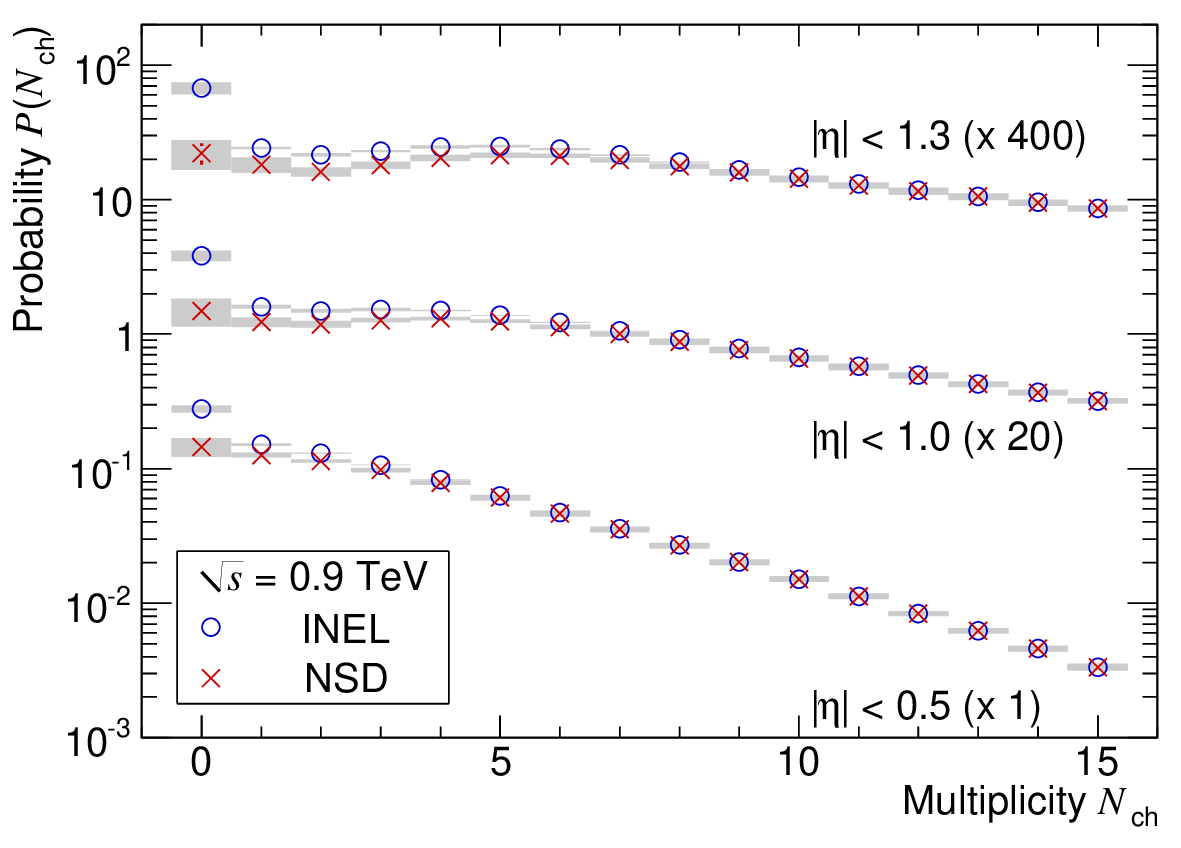}
  \includegraphics[width=\columnwidth]{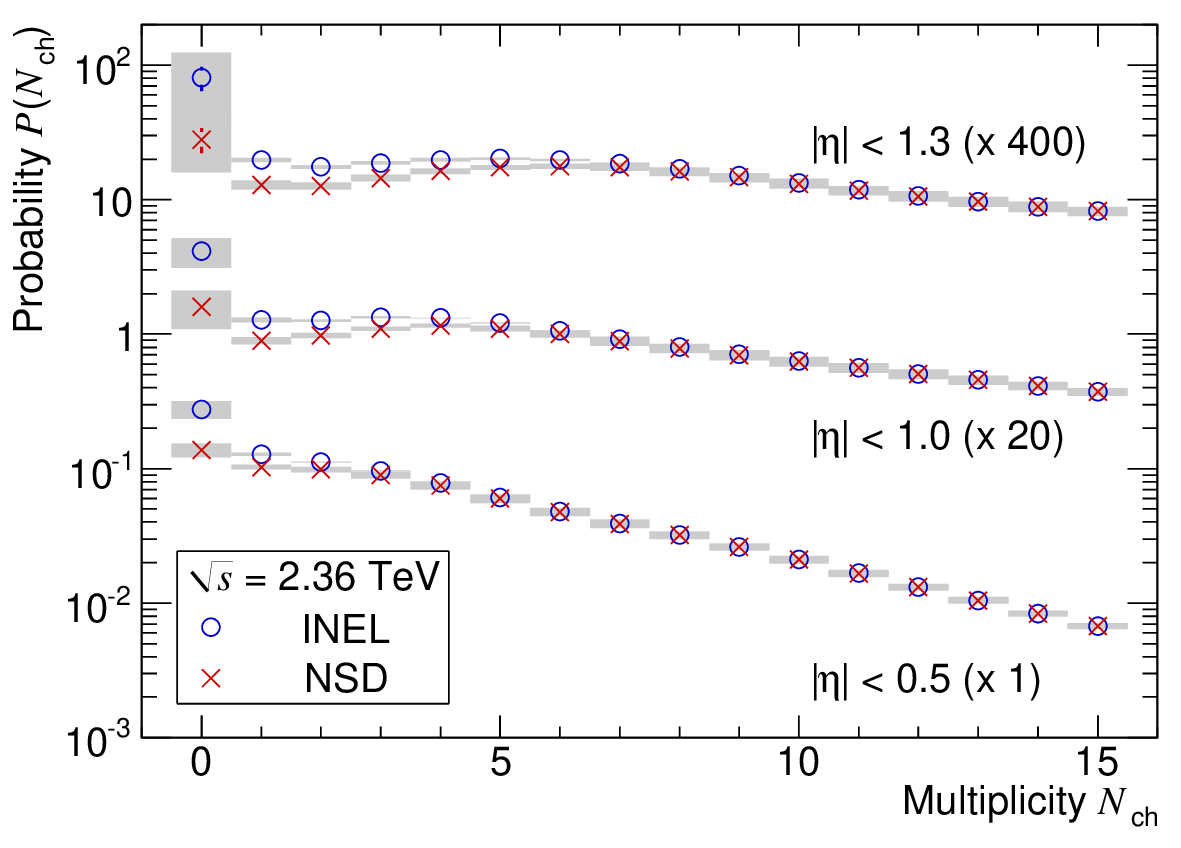}
  \caption{Expanded views of the low-multiplicity region of corrected multiplicity distributions for INEL and NSD events, left for 0.9~TeV and right for 2.36~TeV data. The gray bands indicate the systematic uncertainty. Distribution for NSD events are not normalized to unity but scaled down in such a way that the distributions for INEL and NSD events match at high multiplicities, which makes the difference at low multiplicity clearly visible. Left: data at $\cms = 0.9$~TeV. Right: data at $\cms = 2.36$~TeV. Note that for \etain{1.0} and \etain{1.3} the distributions have been scaled for clarity by the factor indicated.}
  \label{figure_true}
\end{figure*}

The multiplicity distributions of charged particles were measured in three pseudorapidity intervals at both energies. These distributions, corrected as described above, are shown in Fig.~\ref{figure_ua5} (left) and Fig.~\ref{figure_ua5} (right) respectively, for $\cms = 0.9$~TeV and $\cms = 2.36$~TeV for NSD events.
The difference between the multiplicity distributions for NSD and for INEL events only becomes significant at low multiplicities (see Fig.~\ref{figure_true}), as expected.

\begin{figure*}[htb]
\centering
  \includegraphics[width=\columnwidth]{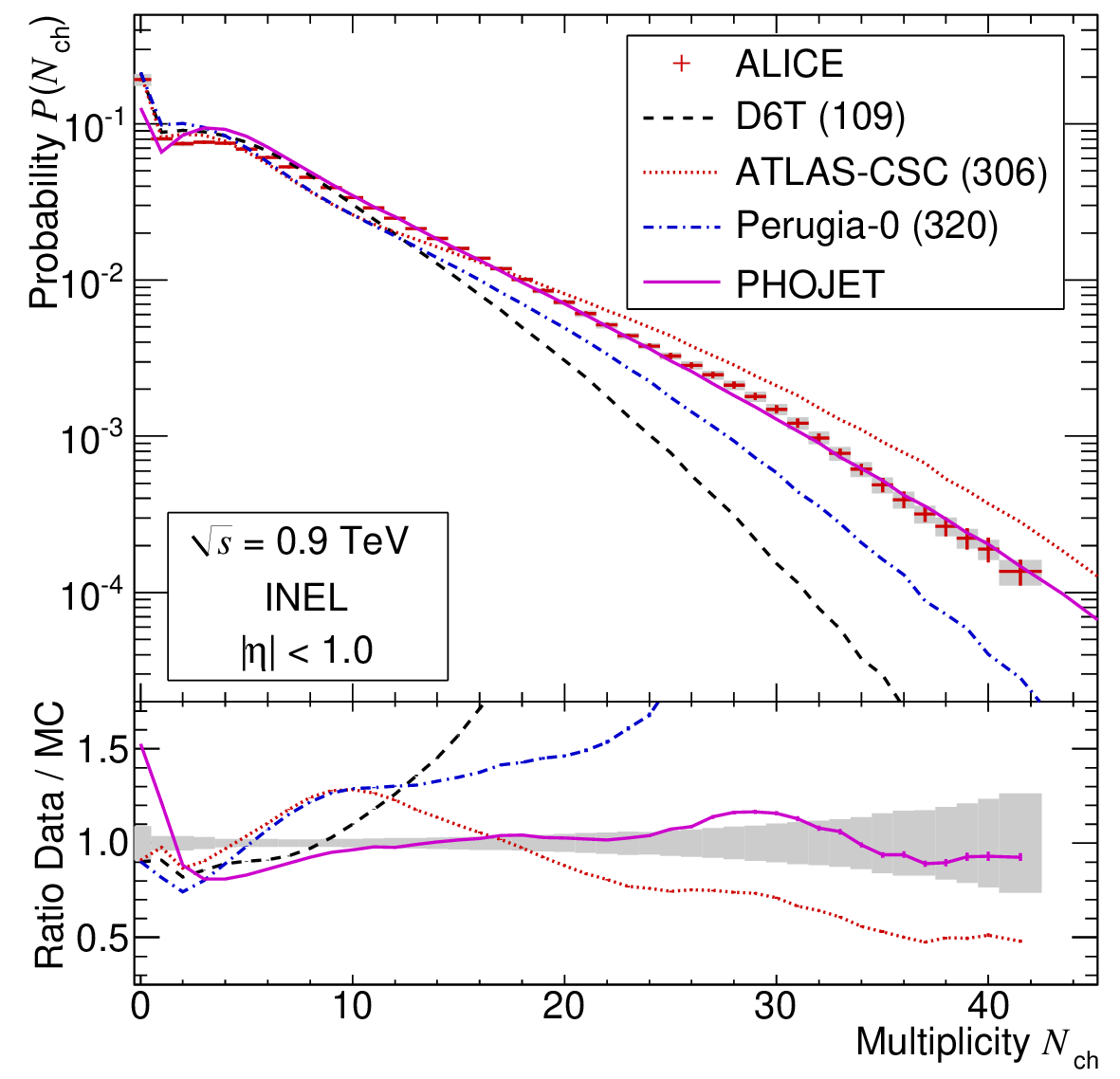}
  \includegraphics[width=\columnwidth]{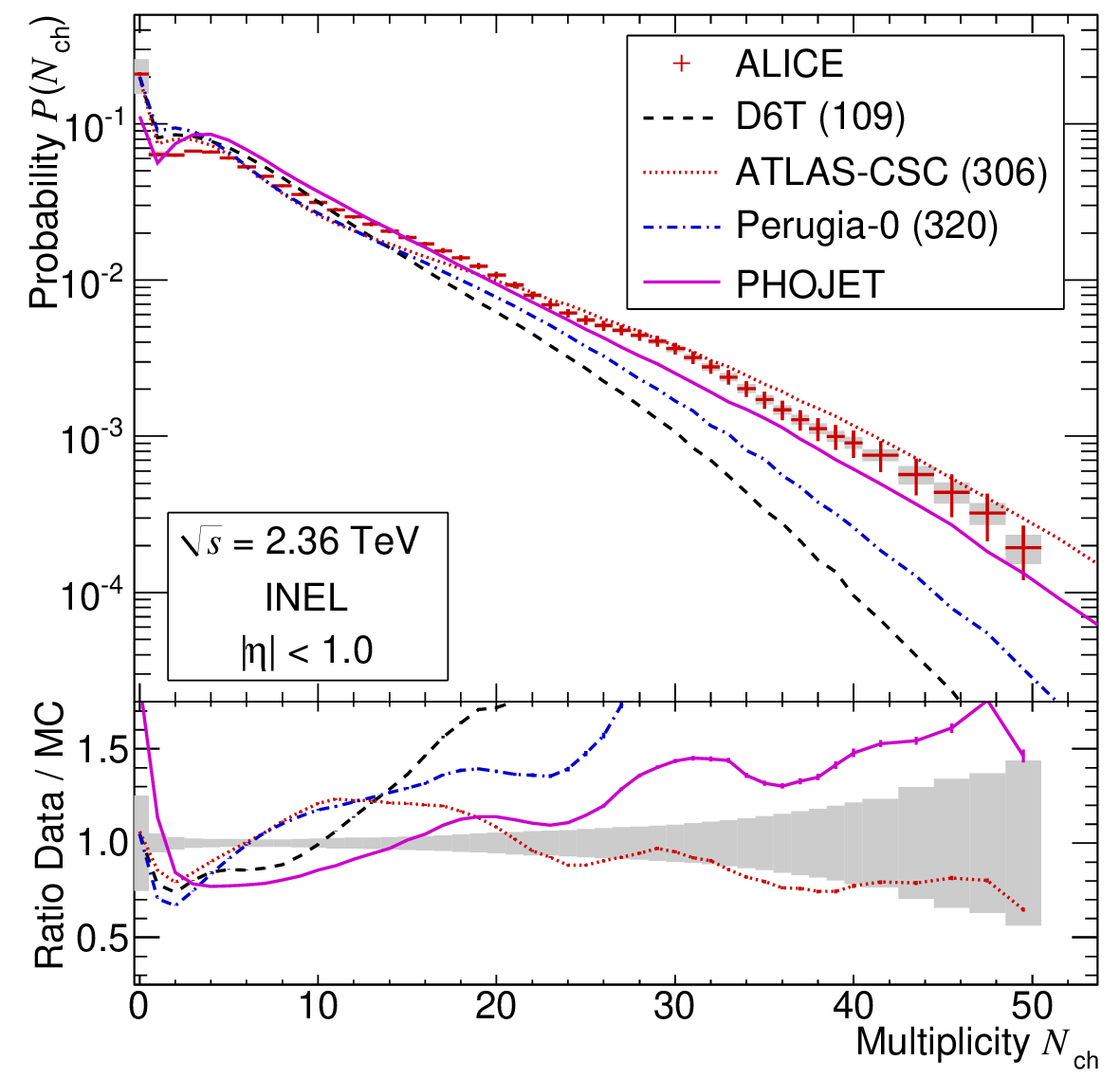}
  \caption{Comparison of measured multiplicity distributions for INEL events to models for the pseudorapidity range \etain{1.0}.
  Predictions are shown based on the PHOJET model~\cite{PhoJet} (solid line) and PYTHIA tunes: D6T~\cite{D6Ttune} (dashed line), ATLAS-CSC~\cite{CSCtune} (dotted line), and Perugia-0~\cite{Perugiatune} (dash-dotted line).
  The error bars for data points represent statistical uncertainties, the shaded areas represent systematic uncertainties. Left: data at 0.9~TeV. Right: data at 2.36~TeV. For both cases the ratios between the measured values and model calculations are shown in the lower part with the same convention. The shaded areas represent the combined statistical and systematic uncertainties.}
  \label{modcomp}
\end{figure*}

In the two larger pseudorapidity intervals, small wavy fluctuations are seen at multiplicities above 25. While visually they may appear to be significant, one should note that the errors in the deconvoluted distribution are correlated over a range comparable to the multiplicity resolution (see Fig.~\ref{figure_responsematrix}). We studied the significance of these fluctuations assuming an exponential shape of the corrected distribution in the corresponding multiplicity range. Applying the response matrix to this smooth distribution and comparing with the measured raw distribution, we find differences of up to two standard deviations in some of the corresponding raw data bins. Therefore, we conclude that while the structures are related to fluctuations in the raw data, they are not significant, and that the uncertainty bands should be seen as one-standard-deviation envelopes of the deconvoluted distributions. Similar observations for a different deconvolution method were made by UA5 in~\cite{ua5_mult3}.

The multiplicity distributions were fitted with a Nega\-tive-Binomial Distribution (NBD) and at both energies satisfactory descriptions were obtained, as shown in  Fig.~\ref{figure_ua5}. Fitting the spectra with the sum of two NBDs, as suggested in~\cite{ua5binom}, did not significantly improve the description of the data.

\begin{table*}[t]
\centering
  \caption{Mean multiplicity and $C_q$-moments (\ref{Cq}) of the multiplicity distributions measured by UA5~\cite{ua5_mult3} in proton--antiproton collisions at $\cms = 0.9$~TeV, and by ALICE at $\cms = 0.9$~TeV and $2.36$~TeV, for NSD events in three different pseudorapidity intervals. The first error is statistical and the second systematic.}
  \label{tabmoments}
  \vspace{0.2cm}
  \begin{tabular}{llll}
    \hline
    \tblspc
    & \multicolumn{1}{c}{UA5 p$\pbar$ } & \multicolumn{2}{c}{ALICE pp} \\
    \cline{2-4}
    \tblspc
    & \multicolumn{2}{c}{$\cms = 0.9$~TeV} & \multicolumn{1}{c}{$\cms = 2.36$~TeV}\\
    \hline
    \multicolumn{4}{c}{\tblspc\etain{0.5}} \\
    \hline
    \tblspc
$\langle N_\mathrm{ch} \rangle$ & $3.61 \pm 0.04 \pm 0.12$    &    $3.60 \pm 0.02 \pm 0.11$     &  $4.47 \pm 0.03 \pm 0.10$    \\
$C_2$   & $1.94 \pm 0.02 \pm 0.04 $ & $ 1.96 \pm 0.01 \pm 0.06 $ & $2.02 \pm 0.01 \pm 0.04$\\
$C_3$   & $ 5.4 \pm 0.2 \pm 0.3 $   & $ 5.35 \pm 0.06 \pm 0.31 $ & $5.76 \pm 0.09 \pm 0.26$\\
$C_4$   & $  19 \pm 1  \pm 1    $   & $ 18.3 \pm 0.4 \pm 1.6 $ & $20.6 \pm 0.6 \pm 1.4$\\
    \hline
    \multicolumn{4}{c}{\tblspc\etain{1.0}} \\
    \hline
    \tblspc
$\langle N_\mathrm{ch} \rangle$ & $7.38 \pm 0.08 \pm 0.27$ & $7.38 \pm 0.03 \pm 0.17$ & $9.08 \pm 0.06 \pm 0.29$ \\
$C_2$   & $1.75 \pm 0.02 \pm 0.04 $  & $1.77 \pm 0.01 \pm  0.04$ & $1.84 \pm 0.01 \pm 0.06$\\
$C_3$   & $4.4  \pm 0.1  \pm 0.1  $  & $4.25 \pm 0.03 \pm  0.20$ & $4.65 \pm 0.06 \pm 0.30$\\
$C_4$   & $ 14.1\pm 0.9  \pm 1.2  $  & $12.6 \pm 0.1 \pm  0.9$ & $14.3 \pm 0.3 \pm 1.4$\\
  \hline
  \multicolumn{4}{c}{\tblspc\etain{1.3}} \\
    \hline
    \tblspc
$\langle N_\mathrm{ch} \rangle$ && $9.73 \pm 0.12 \pm 0.19$ & $11.86 \pm 0.22 \pm 0.45$ \\
$C_2$   &  & $1.70 \pm 0.02 \pm 0.03 $  & $1.79 \pm 0.03 \pm 0.07$ \\
$C_3$   &  & $3.91 \pm 0.10 \pm 0.15  $ & $4.35 \pm 0.16 \pm 0.33$ \\
$C_4$   &  & $10.9\pm 0.4  \pm 0.6  $ & $12.8 \pm 0.7 \pm 1.5$ \\
  \hline
  \end{tabular}
\end{table*}

A comparison of the data to the multiplicity distributions obtained with the event generators is shown in Fig.~\ref{modcomp} for \etain{1.0}.
At low multiplicities ($< 20$) discrepancies are observed at both energies and for all models. At high multiplicities
 and for the 0.9~TeV sample, the PHOJET model agrees well with the data. The PYTHIA tunes D6T and Perugia-0 underestimate the data at high multiplicities and the ATLAS-CSC tune is above the data in this region. At 2.36~TeV, ATLAS-CSC tune of PYTHIA and, to some extent, PHOJET are close to the data. The ratios of data over Monte Carlo calculations are very similar in all three pseudorapidity ranges and suggests that the stronger rise with energy seen in the charged-particle density is, at least partly, due to a larger fraction of high-multiplicity events.

\bfig
  \includegraphics[width=\columnwidth]{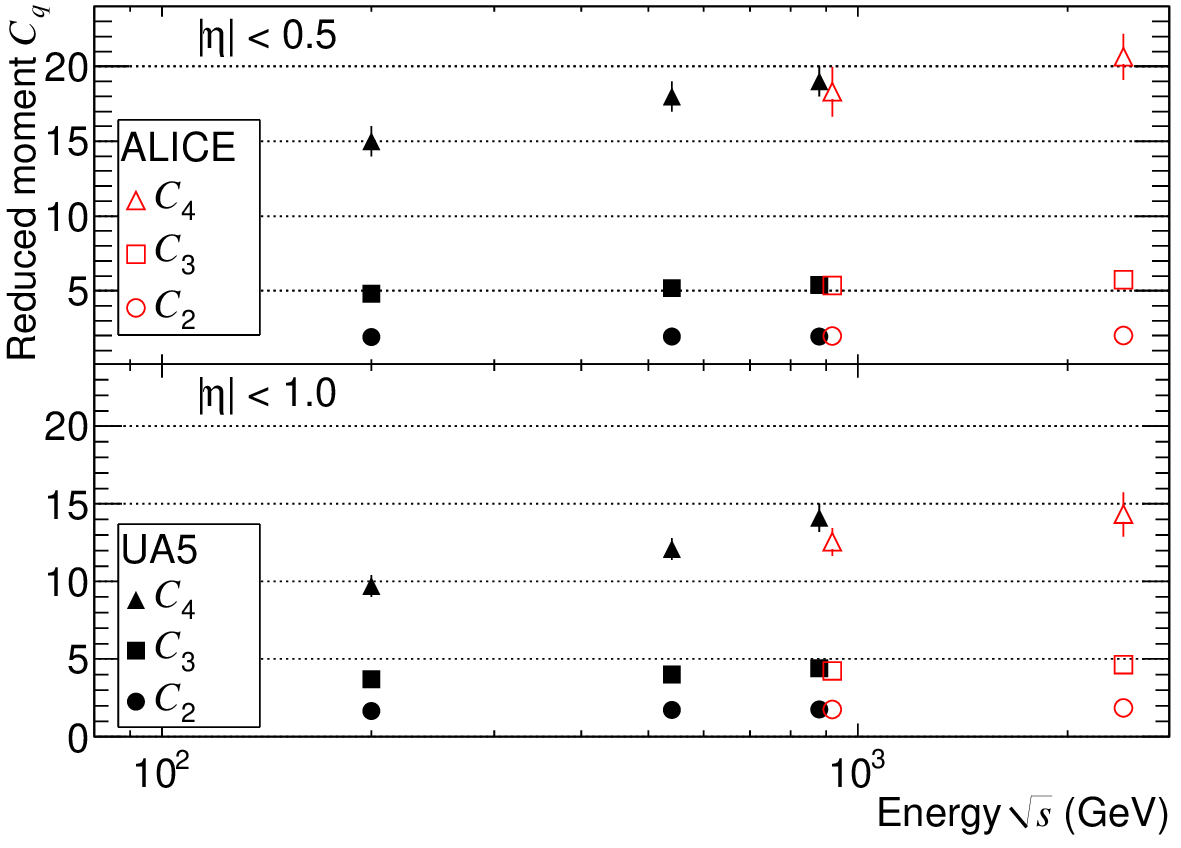}
  \caption{Energy dependence of the $C_q$-moments (\ref{Cq}) of the multiplicity distributions measured by UA5~\cite{ua5_mult3} and ALICE at both energies for NSD events in two different pseudorapidity intervals. The error bars represent the combined statistical and systematic uncertainties. The data at \unit[0.9]{TeV} are displaced horizontally for visibility.}
  \label{figmom}
\efig

\bfig
  \includegraphics[width=\columnwidth]{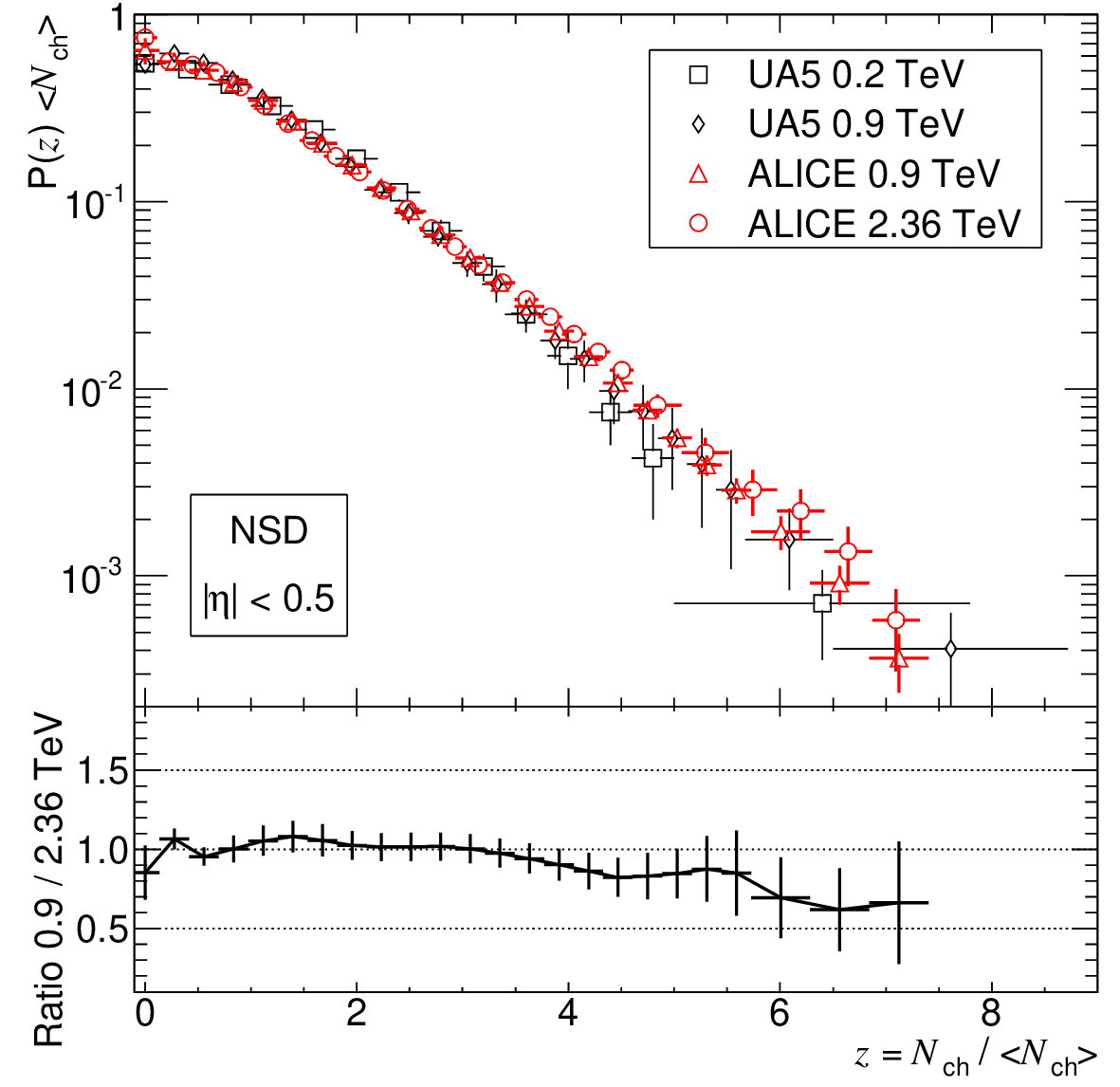}
  \caption{Comparison of multiplicity distributions in KNO variables measured by UA5~\cite{binom,ua5_mult3} in proton--antiproton collisions at $\cms = 0.2$~TeV and $0.9$~TeV, and by ALICE at $\cms = 0.9$~TeV and $2.36$~TeV, for NSD events in \etain{0.5}. In the lower part the ratio between ALICE measurements at 0.9~TeV and 2.36~TeV is shown. The error bars represent the combined statistical and systematic uncertainties.}
  \label{figKNO}
\efig

From these multiplicity distributions we have calculated the mean multiplicity and first reduced moments
\begin{equation}
C_q \equiv \langle N_\mathrm{ch}^q \rangle / \langle N_\mathrm{ch} \rangle^q ,
\label{Cq}
\end{equation}
summarized in Table~\ref{tabmoments}. For \etain{0.5} and \etain{1.0} our results are compared to the UA5 measurement for p$\pbar$ collisions at $\cms = 0.9$~TeV~\cite{ua5_mult3}. Note that the mean multiplicities quoted in this table are those calculated from the multiplicity distributions and are therefore slightly different from the values given in Table~\ref{tabdNdeta}. The value of the pseudorapidity density obtained when averaging the multiplicity distribution for \etain{0.5} is consistent with the value obtained in the pseudorapidity-density analysis. This is an important consistency check, since the correction methods in the pseudorapidity-density and multiplicity-distribution analyses are different.

Our data are consistent with UA5 proton--antiproton measurements at \unit[900]{GeV} (Fig.~\ref{figure_ua5}a and Table~\ref{tabmoments}).
The energy dependence of the reduced moments $C_q$, shown in Fig.~\ref{figmom}, indicates a slight increase, which is not significant given the size of our systematic uncertainties. Systematic uncertainties are assumed to be uncorrelated between energies.
A similar conclusion about the shape evolution of multiplicity distributions can be drawn from Fig.~\ref{figKNO}, where we compare our measurements, plotted in terms of KNO variables, at the two energies and
UA5 p$\pbar$ data at $\cms = 0.2$ and 0.9~TeV, for NSD collisions and pseudorapidity interval \etain{0.5}. While KNO scaling gives a reasonable description of the data from 0.2 to 2.36~TeV, the ratio between the 0.9~TeV  and 2.36~TeV data shows a slight departure from unity above $z = 4$.

\section{Conclusion}

 We report high-statistics measurements of the charged-primary particle pseudorapidity density and multiplicity distributions in proton--proton collisions at centre-of-mass energies of $0.9$~TeV and 2.36~TeV with the ALICE detector. The results at 0.9~TeV are consistent with  UA5 p$\pbar$ measurements at the same energy. At both energies, our data are consistent with the CMS measurement, and compared to various models for which they provide further constraints.
 None of the investigated models and tunes describes the average multiplicities and the multiplicity distributions well. In particular, they underestimate  the increase in the average multiplicity seen in the data between 0.9~TeV and 2.36~TeV. At 0.9~TeV, the high-multiplicity tail of the distributions is best described by the PHOJET model, while at 2.36~TeV, PYTHIA tune ATLAS-CSC is closest to the data.

The multiplicity distributions at both energies and in pseudorapidity ranges up to \etain{1.3} are described well with negative binomial distributions. The shape evolution of the multiplicity distributions with energy was studied in terms of KNO-scaling variables, and by extracting reduced moments of the distributions. A slight, but only marginally significant evolution in the shape is visible in the data for $z > 4$, possibly indicating an increasing fraction of events with the highest multiplicity.
This issue will be studied further using the data collected from forthcoming higher-energy runs at the LHC.

%% file: acknowledgements3.tex
\begin{acknowledgement}

\section*{Acknowledgements}

The ALICE collaboration would like to thank all its engineers and technicians for their invaluable contributions to the construction of the experiment and the CERN accelerator teams for the outstanding performance of the LHC complex.

The ALICE collaboration acknowledges the following funding agencies for their support in building and
running the ALICE detector:
\begin{itemize}
\item{}
Calouste Gulbenkian Foundation from Lisbon and Swiss Fonds Kidagan, Armenia;
\item{}
Conselho Nacional de Desenvolvimento Cient\'{\i}fico e Tecnol\'{o}gico (CNPq), Financiadora de Estudos e Projetos (FI\-N\-EP),
Funda\c{c}\~{a}o de Amparo \`{a} Pesquisa do Estado de S\~{a}o Paulo (FAPESP);
\item{}
National Natural Science Foundation of China (NSFC), the Chinese Ministry of Education (CMOE)
and the Ministry of Science and Technology of China (MSTC);
\item{}
Ministry of Education and Youth of the Czech Republic;
\item{}
Danish Natural Science Research Council, the Carlsberg Foundation and the Danish National Research Foundation;
\item{}
The European Research Council under the European Community's Seventh Framework Programme;
\item{}
Helsinki Institute of Physics and the Academy of Finland;
\item{}
French CNRS-IN2P3, the `Region Pays de Loire', `Region Alsace', `Region Auvergne' and CEA, France;
\item{}
German BMBF and the Helmholtz Association;
\item{}
Hungarian OTKA and National Office for Research and Technology (NKTH);
\item{}
Department of Atomic Energy and Department of Science and Technology of the Government of India;
\item{}
Istituto Nazionale di Fisica Nucleare (INFN) of Italy;
\item{}
MEXT Grant-in-Aid for Specially Promoted Research, Ja\-pan;
\item{}
Joint Institute for Nuclear Research, Dubna;
\item{}
Korea Foundation for International Cooperation of Science and Technology (KICOS);
\item{}
CONACYT, DGAPA, M\'{e}xico, ALFA-EC and the HELEN Program (High-Energy physics Latin-American--European Network);
\item{}
Stichting voor Fundamenteel Onderzoek der Materie (FOM) and the Nederlandse Organisatie voor Wetenschappelijk Onderzoek (NWO), Netherlands;
\item{}
Research Council of Norway (NFR);
\item{}
Polish Ministry of Science and Higher Education;
\item{}
National Authority for Scientific Research - NASR (Autontatea Nationala pentru Cercetare Stiintifica - ANCS);
\item{}
Federal Agency of Science of the Ministry of Education and Science of Russian Federation, International Science and
Technology Center, Russian Academy of Sciences, Russian Federal Agency of Atomic Energy, Russian Federal Agency for Science and Innovations and CERN-INTAS;
\item{}
Ministry of Education of Slovakia;
\item{}
CIEMAT, EELA, Ministerio de Educaci\'{o}n y Ciencia of Spain, Xunta de Galicia (Conseller\'{\i}a de Educaci\'{o}n),
CEA\-DEN, Cubaenerg\'{\i}a, Cuba, and IAEA (International Atomic Energy Agency);
\item{}
Swedish Reseach Council (VR) and Knut $\&$ Alice Wallenberg Foundation (KAW);
\item{}
Ukraine Ministry of Education and Science;
\item{}
United Kingdom Science and Technology Facilities Council (STFC);

\item{}
The United States Department of Energy, the United States National
Science Foundation, the State of Texas, and the State of Ohio.

\end{itemize}
\end{acknowledgement}